\documentclass[lettersize,journal]{IEEEtran}
\usepackage{lipsum}
\usepackage{tikz}
\usepackage{amsmath}
\usepackage{stfloats}
\usepackage{url}
\usepackage{verbatim}
\usepackage{graphicx}
\usepackage{dsfont}
\usepackage{amsfonts}
\usepackage{array}
\usepackage{hyperref}
\usepackage[caption=false,font=normalsize,labelfont=sf,textfont=sf]{subfig}
\usepackage{algorithm2e}
\RestyleAlgo{ruled}
\SetKwComment{Comment}{/* }{ */}
\usepackage{booktabs}
\usepackage{filecontents}
\usepackage{multirow}

\usepackage{amssymb}

\newcommand{\mcl}[1]{\mathcal{#1}}
\newcommand{\mbf}[1]{\mathbf{#1}}
\newcommand{\mbb}[1]{\mathbb{#1}}

\newcommand{\x}[0]{\mathbf{x}}

\newcommand{\y}[0]{\mathbf{y}}
\newcommand{\z}[0]{\mathbf{z}}
\newcommand{\s}[0]{\mathbf{s}}

\newcommand{\bd}[0]{\bar{d}}
\newcommand{\bdy}[0]{\bar{d}^y}
\newcommand{\bdyh}[0]{\bar{d}^{\hat{y}}}

\newcommand{\ryh}[0]{r^{\hat{y}}}
\newcommand{\cb}[0]{\mathbf{c}}
\newcommand{\cy}[0]{\mathbf{c}^{y}}

\newcommand{\yh}[0]{\hat{y}}
\newcommand{\yt}[0]{\tilde{y}}
\newcommand{\ybt}[0]{\tilde{\mathbf{y}}}
\newcommand{\ybh}[0]{\hat{\mathbf{y}}}

\newcommand{\sm}[0]{f_{\text{sm}}}

\newcommand{\fy}[0]{\tilde{\y}}

\newcommand{\erfc}[0]{\text{erfc}}
\newcommand{\header}[1]{\vskip 5pt \noindent \textbf{#1}}
\newcommand{\sqs}[0]{\text{sqs}}
\newcommand{\cqs}[0]{\text{cqs}}

\def\BibTeX{{\rm B\kern-.05em{\sc i\kern-.025em b}\kern-.08em
    T\kern-.1667em\lower.7ex\hbox{E}\kern-.125emX}}
\usepackage{balance}

\begin{document}
\title{QUEEN:\\Query Unlearning against Model Extraction}
\author{
    Huajie Chen, Tianqing Zhu$^*$, Lefeng Zhang, Bo Liu, Derui Wang ,Wanlei Zhou, and Minhui Xue
    \thanks{$^*$Tianqing Zhu is the corresponding author.}
    \thanks{Huajie Chen and Bo Liu are with University of Technology Sydney.}
    \thanks{Tianqing Zhu, Lefeng Zhang, and Wanlei Zhou are with City University of Macau.}
    \thanks{Derui Wang and Minhui Xue are with CSIRO Data61.}
}

\markboth{Journal of \LaTeX\ Class Files,~Vol.~18, No.~9, September~2020}%
{How to Use the IEEEtran \LaTeX \ Templates}

\maketitle

\begin{abstract}
    Model extraction attacks currently pose a non-negligible threat to the security and privacy of deep learning models.
    By querying the model with a small dataset and using the query results as the ground-truth labels, an adversary can steal a piracy model with performance comparable to the original model.
    Two key issues that cause the threat are, on the one hand, accurate and unlimited queries can be obtained by the adversary; 
    on the other hand, the adversary can aggregate the query results to train the model step by step. 
    The existing defenses usually employ model watermarking or fingerprinting to protect the ownership.
    However, these methods cannot proactively prevent the violation from happening.
    To mitigate the threat, we propose QUEEN (QUEry unlEarNing) that proactively launches counterattacks on potential model extraction attacks from the very beginning.
    To limit the potential threat, QUEEN has sensitivity measurement and outputs perturbation that prevents the adversary from training a piracy model with high performance. 
    In sensitivity measurement, QUEEN measures the single query sensitivity by its distance from the center of its cluster in the feature space.
    To reduce the learning accuracy of attacks, for the highly sensitive query batch, QUEEN applies query unlearning, which is implemented by gradient reverse to perturb the softmax output such that the piracy model will generate reverse gradients to worsen its performance unconsciously.
    Experiments show that QUEEN outperforms the state-of-the-art defenses against various model extraction attacks with a relatively low cost to the model accuracy. The artifact is publicly available at \url{https://anonymous.4open.science/r/queen\_implementation-5408/}.
\end{abstract}


\section{Introduction}
Having achieved revolutionary breakthroughs in various domains, deep neural networks (DNNs) are currently being employed in diverse areas to solve sophisticated real-world problems. 
The cost of training a high-performance DNN is non-negligible due to the high cost of the large volume of dataset collection, long training time, hardware consumption, etc. 
For instance, Microsoft has spent $1$ billion dollars on OpenAI to develop a large language deep neural network models \cite{gozalo2023chatgpt}. 
Therefore, the deep learning models are considered valuable intellectual properties to be protected by the model owners (referred to as \textit{defenders} in the following).
Nowadays, \textit{Model Extraction Attack} (MEA)\cite{chen2023ddae} is considered one of the most critical threats to DNN properties. 
In MEA, an adversary can establish a piracy model that has the same functionality and comparable performance as that of the protectee model by sending queries to the original models. 
This type of attack can lead to serious consequences in terms of copyrights as the adversary can steal from real-world Machine Learning as a Service (MLaaS) models at a very low cost. 
Meanwhile, MEAs are often used as the initial step for other attacks such as membership inference or model inversion attacks, where the adversary trains shadow models in the manner of MEAs \cite{choquette-choolabelonlymembership,liu2022membership}. 
Hence, we urgently need a series of defenses to counter MEAs, especially in the era of large model.

\begin{figure}[t]
    \centering
    \includegraphics[width=.4\textwidth]{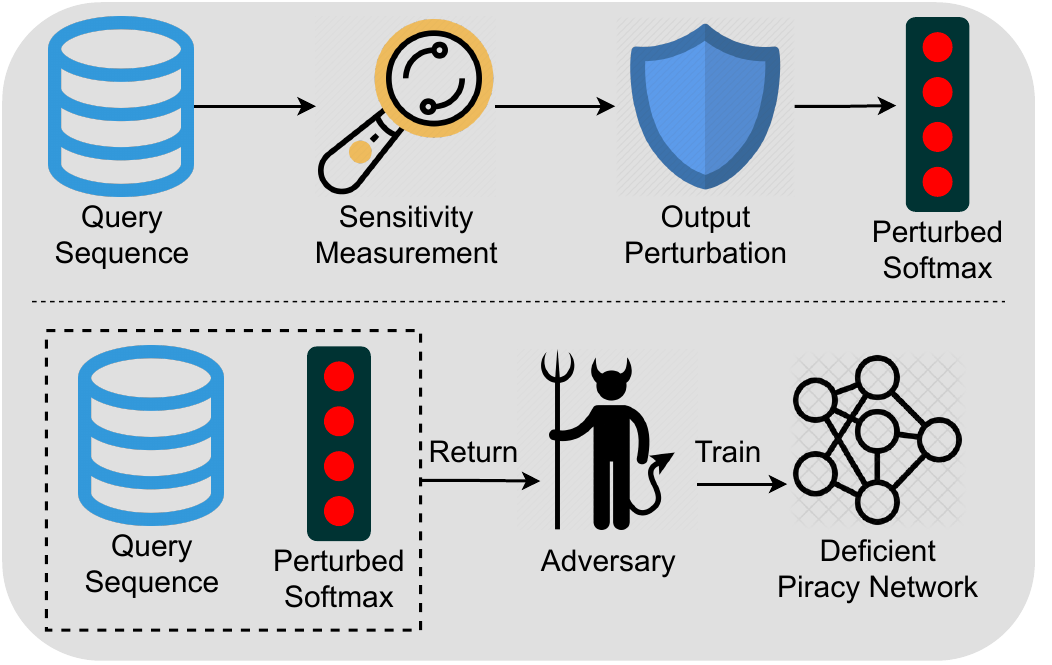}
    \caption{The overview of QUEEN. For any sequence of queries, QUEEN first measures the sensitivity of each query, and then performs output perturbation accordingly to create perturbed softmax outputs. The piracy network trained with such queries and perturbed outputs will not have comparable performance to the original model.}
    \label{fig:queen_overview}
\end{figure}

To mitigate the threat of MEAs, extensive efforts have been made to enhance defense strategies. 
The existing defense methods can be classified into two classes, passive and active defenses. 
Passive defenses mainly comprise model watermarking and fingerprinting, where the defender extracts the embedded watermark from the piracy model or uses the fingerprints to get pre-defined output as proof of copyright violation~\cite{cong2022sslguard,jia2021entangled,lukas2019deep,peng2022fingerprinting,guan2022you}.
Despite the effectiveness of the passive defenses, the validation can only be conducted \textit{after the violation}, and the adversary could have already benefited from the piracy model before the crime is discovered. 
Moreover, if the adversary uses the piracy model only as a proxy to launch further attacks, or the adversary keeps the piracy model for private usage, the defender can no longer protect the protectee model with the passive defenses.


In contrast, active defenses \cite{DziedzicIncreasingTheCost, lee2018defending, kariyappa2020defending} proactively prevent MEAs from happening by jeopardizing the training process with deliberately falsified outputs or lengthening the query time.
Active defenses decrease the performance of the piracy models by perturbing the outputs of the protectee model, resulting in a degraded performance of the protectee model.
Additionally, the majority of active defenses perturb the outputs randomly, which degrades the protectee model's performance significantly.
Alternatively, the defender can also extend the query time of the adversary.
The current MLaaS provides most services in a black-box setting, where the user can only query the model and receive the query outputs. 
Hence, we aim to address two questions to counter the MEAs: 
\textit{1) Can we detect threats from the queries? 
2) Can we proactively eliminate the threat before the violation occurs?}

We have explored the possibilities to answer above questions. For the first question, we find that the threat queries show potential trend to cover all classes in the DNN model. 
In other words, the coverage of those threaten queries is much broader than normal queries series. 
This enlightens us to estimate the coverage of the query set to detect threats. 
For the second question, when we identify the threat, it is possible for defender to provide some perturbed query answers that can mislead the shadow models, so it is possible to eliminate the threat before it actually happens. 

Due to the above positive rationale, we propose QUEEN (QUEry unlEarNing), which proactively launches counterattacks on potential security risks accompanied by sensitivity measurement. As depicted in Figure \ref{fig:queen_overview}, there are two main components in QUEEN, namely \textbf{sensitivity measurement} and \textbf{output perturbation}. Given an arbitrary query sequence, the defender estimates the risk of the query by computing the single query sensitivity (SQS) and iteratively updating the cumulative query sensitivity (CQS) class-wise. In output perturbation, if the CQS of the predicted class of the query does not exceed the pre-defined threshold, the defender will perturb the query in the feature space to obfuscate the softmax output. If the CQS exceeds the threshold, the defender launches gradient reverse to make the piracy model generate gradients that worsen the model during training.

Several challenges are identified and resolved during our research: 
\textit{1) How to effectively measure the sensitivity of single queries?} 
For each class in the training dataset, the defender extracts the features of the training samples, and defines the cluster center of each class as the most sensitive point. The distance between the feature of the query and the cluster center is used to estimate the SQS. To estimate the CQS, each query feature is treated as a hypersphere. By computing the ratio of the query features' volume to the volume of the sensitive region, the defender can update the CQS by class query-wise. 
\textit{2) How to guarantee that the piracy model produces reverse gradients when trained using the falsified confidence vector?} 
To launch gradient reverse, the defender must estimate the output of the piracy model given an input. 
Thus, the defender trains a set of shadow models using subsets of the training dataset to represent the incomplete piracy model. 
To increase the randomness of the results, a fraction of the shadow models are randomly drawn from the set to produce the average confidence vector for each query. 
With the estimated piracy confidence vector and that of the protectee model, the defender is able to launch gradient reverse.


Our contributions are summarized as follows.
\begin{itemize}
    \item We propose a novel counterattack against model extraction named QUEEN that proactively sabotages the MEAs before the copyright violation happens. QUEEN can be efficiently integrated with generic classification networks without interfering with the training process.
    
    \item We have designed novel sensitivity measurement and output perturbation algorithms that can efficiently and precisely prevent MEAs while maintaining the prediction accuracy of the protectee model. The output is selectively perturbed to worsen the piracy model.

    \item We have conducted extensive experiments to validate the effectiveness of QUEEN, where QUEEN has outperformed the SOTA defenses.
\end{itemize}


\section{Preliminaries and Related Work}
The notations are listed in Table \ref{tab:notations}.

\begin{table}[t]\caption{Summary of notations.}
    \label{tab:notations}
    \centering
    \begin{tabular}{c|l}
    \toprule
         Symbols & Definitions\\
         \midrule
         $\mcl D, \mcl F$ & The problem domain/feature space\\
         $D, D'$ & The original/auxiliary dataset\\
         $\x, \x'$ & The original/auxiliary data sample\\
         $\y, \ybh, \ybt$ & The original/predicted/perturbed confidence vector\\
         $y, \yh, \yt$ & The original/predicted/perturbed label\\
         $f, h$ & The protectee/piracy model\\
         $\theta, \tau$ & The parameters of the protectee/piracy model\\
         $n, m$ & The number of samples in $D/D'$\\
         $f^E, f^C$ & The feature extraction/classification block of $f$\\
         $U, Z$ & The feature set in $\mcl F/\mbb R^2$\\
         \bottomrule
    \end{tabular}
\end{table}

\subsection{Model Extraction Attacks}
Given the protectee model $f(\cdot; \theta): \mbb R^m \mapsto \mbb R^n$ parameterized by $\theta$, the objective of MEAs is to establish a piracy model $h(\cdot; \tau): \mbb R^m \mapsto \mbb R^n$ parameterized by $\tau$, where $h$ mimics the functionality of $f$ by optimizing
\begin{equation}
    \begin{aligned}
        \underset{\tau}{\arg \min} \ \mcl L(f(\x; \theta), h(\x; \tau)),
    \end{aligned}
\end{equation}
where $\mcl L$ and $\x \in \mbb R^m$ respectively denote the loss function and the query sample. The adversary has only black-box access to $f$, where the adversary has zero knowledge about the architecture, parameters or hyperparameters of $f$. Normally, the adversary is assumed to have a limited number of unlabelled data samples and computational resources, otherwise the adversary can train the model independently. 
The dataset owned by the adversary can either be in-distribution or out-of-distribution to the training dataset of $f$. 
The auxiliary dataset can be collected online.


Querying the protectee model with natural samples is the most direct attack. However, due to the limited query budget, the adversary tends to select the optimal samples for querying. 
For instance, semi-supervised learning \cite{correia2018copycat}, reinforcement learning \cite{orekondy2019knockoff}, and active learning \cite{pal2020activethief} are used for query sampling selection. 
Additionally, using synthetic data to query the protectee model can also achieve considerable attack accuracy. 
There are various ways to generate the synthetic samples, such as FGSM \cite{papernot2017practical}, C\&W attack and feature adversary attack \cite{yu2020cloudleak}. 
For example, FGSM uses the gradient to modify the sample $x$, such that the modified sample deviates from the predicted class given by $h$, which is defined as
\begin{equation}
\begin{aligned}
    x \gets x - \eta \nabla_x \mcl L(x, h(x; \tau)),
\end{aligned}
\end{equation}
where $\eta$ controls the strength of the modification.
Furthermore, even without an auxiliary dataset, the adversary can use generative models to launch data-free MEAs \cite{truong2021datafree,kariyappa2021maze}, but this requires an enormous query budget.

\subsection{Defenses against Model Extraction Attacks}

Generally, the current defenses against MEAs can be divided into passive and active defenses. 

Passive defenses mainly include model watermarking and fingerprinting, where the defender can determine whether a suspect model is a piracy model or not by validation. Model watermarking~\cite{cong2022sslguard,jia2021entangled, li2023black,zhang2023categorical,peng2023you} aims to embed watermarks into the protectee model during training, where a set of trigger samples can make the protectee model generate the pre-defined outputs. 
Model fingerprinting \cite{peng2022fingerprinting, hu2023veridip, lukas2019deep}, in contrast, does not interfere with the training process, and it creates adversarial samples as fingerprints to make the piracy model and the protectee model share the same outputs when given the fingerprints. 
In summary, passive defenses allow the defender to claim the copyright of the piracy model by ownership verification. However, this does not prevent the violation from the root.

Active defenses proactively prevent MEAs by perturbing the outputs of the protectee model or intentionally increasing the query time when the query number ascends. Orekondy et al. \cite{orekondy2019prediction} perturb the output by maximizing the angle deviation between the original and the perturbed gradients. 
Lee et al. \cite{lee2018defending} attach an additional layer to the end of the protectee model to produce random softmax output while maintaining the argmax of the output unchanged. Juuti et al. 
\cite{juuti2019prada} use the Shapiro-Wilk statistics test to distinguish the benign queries from the malicious queries, where the benign samples are believed to fit normal distributions and the malicious do not. Kariyappa et al. \cite{kariyappa2020defending} train a out-of-distribution (OOD) sample detector to differentiate OOD query samples from normal query samples. The query results of OOD samples are perturbed by a misinformation function. Kariyappa et al. \cite{kariyappa2020protecting} train an ensemble of models to detect OOD query samples and perturb the corresponding outputs. 
Zhang et al. \cite{zhang2023apmsa} perturb the query sample to the edge of the decision boundary while keeping the argmax of the output unchanged to produce obfuscated outputs. 
Dziedzic et al. \cite{DziedzicIncreasingTheCost} utilize private aggregation of teacher ensembles (PATE) to measure the privacy risk of each query, and increase the query time when the query number grows by forcing the adversary to solve the proof-of-work (PoW) puzzles.


\section{Problem Definition}
\label{sect:prob_def}

First, we formally define the protectee model and the piracy model. 
A problem domain is denoted by $\mcl D \subseteq \mbb R^M$, where each element $\mbf x \in \mcl D$ is labeled by one of $N$ classes. 
We use $\mbf y \in \mbb R^N$ to denote the one-hot encoded label vector. 
A deep learning model is a function $f(\cdot; \theta): \mbb R^M \mapsto \mbb R^N$, parameterized by $\theta$. 

A protectee model is a deep learning model $f(\cdot; \theta): \mbb R^M \mapsto \mbb R^N$ with $\theta$ as its parameters trained by a defender who owns a dataset $D \subseteq \mcl D$.
$D$ comprises $\{(\mbf x_1, \mbf y_1),...,(\mbf x_n, \mbf y_n)\}$.
The defender trains the protectee model $f$ by taking gradient steps to optimize $\theta$ on
\begin{equation}
    \begin{aligned}
        \nabla_\theta \frac{1}{n} \sum_{i=1}^n \mcl L\bigg( \sm (f(\mbf x_i; \theta)), \mbf y_i \bigg),
    \end{aligned}
\end{equation}
where $\mcl L(\cdot, \cdot)$ denotes the loss function such as the cross entropy (CE) loss or Kullback-Leibler divergence (KLDiv) loss;
$\sm(\cdot)$ denotes the softmax function.

A piracy model is another deep learning model $h(\cdot; \tau): \mbb R^M \mapsto \mbb R^N$ with $\tau$ as its parameters established by the adversary by launching model extraction attack on the protectee model $f$.
$h$ shares the same or similar functionalities with $f$.
The adversary collects an auxiliary dataset $D' \subseteq \mcl D \backslash D$. This means that $D' = \{\mbf x'_1,...,\mbf x'_m\}$ comes from the same problem domain $\mcl D$ but does not overlap with $D$.
Additionally, $\x' \in D'$ is not labeled by the adversary.
The adversary feeds $\mbf x'$ into the protectee model $f$ so as to get the softmax output, i.e., the confidence vector $f(\mbf x')$.
Similarly, the adversary trains the piracy model $h_\tau$ taking gradient steps to optimize $\tau$ on
\begin{equation}
    \begin{aligned}
        \nabla_\tau \frac{1}{m} \sum_i^m \mcl L\bigg( \sm (h(\mbf x_i'; \tau)), \sm(f(\mbf x_i'; \theta)) \bigg), \forall \ \x_i' \in D'.
    \end{aligned}
\end{equation}

\subsection{Threat Model}
We consider the scenarios where not only the label of the prediction, but also the entire confidence vector is of the users' concern.
For example, the confidence vector can be further utilized for downstream tasks such as OOD detection \cite{hendrycks2022scaling} or membership inference attack \cite{ye2022enhanced}.
We define the threat model in terms of the capabilities and limitations of the two parties, namely the adversary and the defender.

\header{The Adversary.}
We assume that the adversary is capable of:
\begin{itemize}
    \item \textit{Collecting public data.} 
    The adversary can access public datasets to construct an auxiliary dataset for model extraction. The auxiliary dataset can be either in-distribution or out-of-distribution to the training dataset. For instance, given that the protectee model is trained on MNIST dataset, FEMNIST dataset is in-distribution, and CIFAR-10 dataset is out-of-distribution.
    
    \item \textit{Accessing the protectee model.} 
    The adversary can query the protectee model with any input with black-box access. It means that the adversary is limited to sending queries to the protectee model and receiving the predictions from the protectee model only. The predictions are presented in the form of complete confidence vectors.
    
    \item \textit{Customizing the training algorithm.} 
    The adversary is able to freely select the network architecture, optimization algorithms, loss functions, etc. However, we assume that the adversary is going to select either the CE or KLDiv loss function, because these two are the most commonly used loss functions in terms of training classifiers.
    
    \item \textit{Knowing the specific defense method.} 
    The adversary knows which exact defense method is used, and therefore, the adversary can launch the most threatening attack on the protectee model.
\end{itemize}

We assume that the adversary is limited to:
\begin{itemize}
    \item \textit{Knowledge of the protectee model.} The adversary does not know the model architecture, parameters, or hyperparameters of the protectee model.
    \item \textit{Access to the training data.} The adversary cannot access the original training samples possessed by the defender.
\end{itemize}

\header{The Defender.}
In contrast, the defender is capable of:
\begin{itemize}
    \item \textit{Completely accessing the protectee model.} The defender has full knowledge about the protectee model including its model architecture, parameters and hyperparameters.
    \item \textit{Completely accessing the original dataset.} The defender can access and modify any samples in the original dataset.
    \item \textit{Utilizing the queries.} For any query sent by the user, the defender is allowed to perform any operation on the query, including storing and analyzing the query.
\end{itemize}

Meanwhile, the defender is limited to:
\begin{itemize}
     \item \textit{Knowledge of the piracy model.} The defender does not know the model architecture, parameters, or hyperparameters of the piracy model.
    \item \textit{Knowledge of the specific attack.} The attack method employed by the attacker remains unknown to the defender.
\end{itemize}

The goal of the adversary is to launch model extraction attack on the protectee model so as to obtain a piracy model that performs comparably to the protectee model. Reversely, the defender aims to differentiate the adversaries from the benign users, and proactively jeopardize any possible threat.

\header{The Adaptive Adversary.}
We further consider that the adaptive adversary knows about the defense mechanism of QUEEN.
Given a perturbed confidence vector $\ybt$, The adaptive adversary aims to recover the clean prediction confidence vector $\ybh = R(\ybt)$ using a recovery function $R(\cdot)$.
Thus, adaptive attacks such as D-DAE \cite{chen2023ddae} and pBayes \cite{tangmodelguard} that establishes $R(\cdot)$ pose the greatest threats to QUEEN.
We thus test the performance of QUEEN against these adaptive attacks in the experiment.

\subsection{Concepts of Defense Mechanism}\label{sect:def_mech}

\header{Central Data vs. Peripheral Data.}
Before designing the method, we try to figure out how the classification works in $f$. 
Let $f^E$ and $f^C$ respectively denote the feature extraction block and the classification block in $f$. Then we have
\begin{equation}
    \begin{aligned}
        f(\x) = f^C(f^E(\x)),
    \end{aligned}
\end{equation}
where $f^E$ maps $\x$ into a feature space $\mcl F \subseteq \mbb R^O$, and $f^C$ maps $f^E(\x)$ into $\mbb R^N$ so as to derive the confidence vector. 
Given the training dataset $D$ MNIST \cite{mnist}, we use $f^E$ to extract the feature $f^E(\x_i), \ \forall \ \x_i \in D$. The features are then projected into a 2D space via t-SNE as depicted in Figure \ref{fig:tsne_decision_boundary} in the Supplemental Materials.

Based on our observation, $f^E(\x_i)$s are separated into clearly distinguishable clusters by their label $\y_i$, which are the black dots. 
We then process the test dataset with the same procedure as above, where we use different colors to represent the classes. 
Two interesting phenomena are observed:
\textit{1) the projection of the test data highly overlaps with that of the training data; 
2) the misclassified test data points fall in places that are distant from the center of the clusters. }
Thus, we assume that the center of each cluster in $\mcl F$ represents the most sensitive region in the cluster. 
Thus, the queries that hit the center of the cluster are the most representative.

To justify our above assumption, we thus conduct a pre-experiment to check whether the central data is more representative than the peripheral data. Here, central data refers to the data whose feature is close to its cluster center, whereas peripheral data refers to the data whose feature is distant from its cluster center. In order to rank the data by their distances to their cluster center, we firstly need to define the cluster center.

\textit{Definition 1--Cluster Center:} \textit{Given a feature cluster $\{f^E(\x_1),...,f^E(\x_n)\}$ sharing the same label $\y$, the cluster center is defined as}
\begin{equation}\label{eq:cluster_center}
    \begin{aligned}
        \cy = \frac{1}{n} \sum_i^n f^E(\x_i),
    \end{aligned}
\end{equation}
\textit{where $\cy$ denotes the center of the feature cluster labeled by $y = \arg \max (\y)$.}

The experimental results show that the central data leads to higher performance of the trained classifier network compared to the peripheral data.
This suggests that the query whose feature is closer to the cluster center is more sensitive.
Section \ref{appdendix:peri_vs_central} in the Supplemental Materials shows the details.



\begin{figure}[t!]
    \centering
    \includegraphics[width=.4\textwidth]{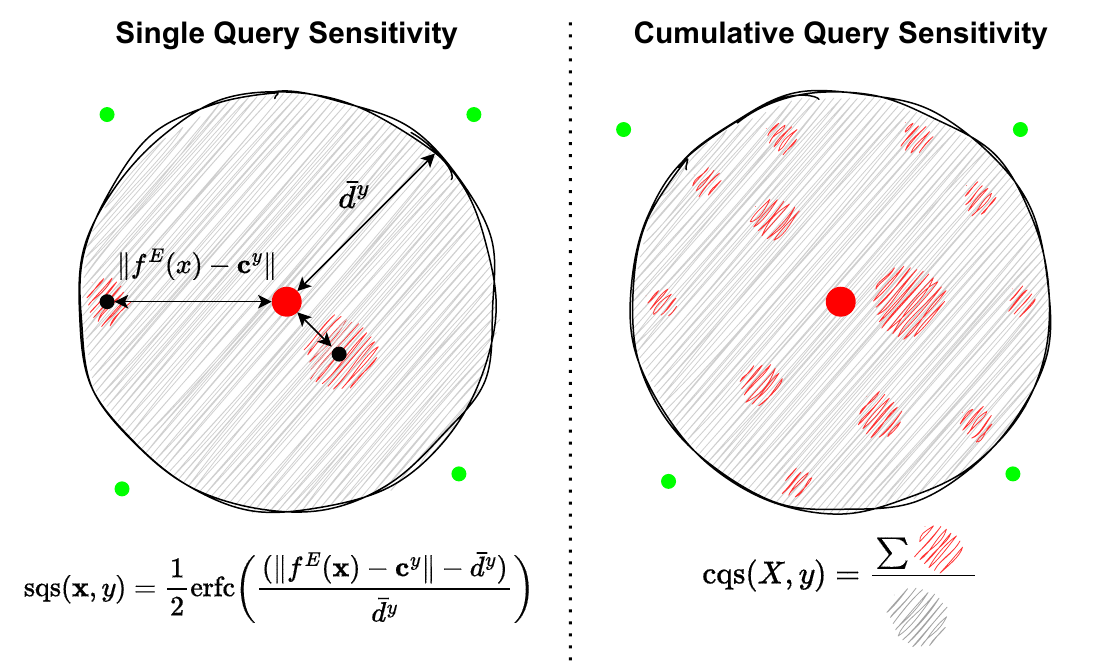}
    \caption{Sensitivity Measurement. The CQS is the sum of the red/queried area over the gray/sensitive area.}
    \label{fig:sensitivity_measurement}
\end{figure}

\header{Single Query Sensitivity.}
Given an arbitrary query, the defender aims to know how sensitive it is when the logit produced by the protectee network is honestly returned. Based on the pre-experiment, we know that the probability of a query being more representative than another gets higher, when its feature is closer to the cluster center than the other. Therefore, we define representative probability (RP) to measure the sensitivity of the query.

\textit{Definition 2--Single Query Sensitivity: Given an arbitrary query $\x$ labeled by $y=\arg\max \y$, the single query sensitivity $sqs(\x, y)$ is measured by the representative probability $P(\x, y)$, which is computed by}
\begin{equation}
    \begin{aligned}
        sqs(\x, y) & = P(\x, y)\\
        & = \frac{1}{2}\erfc \bigg( \frac{ \big( ||f^E(\x) - \cy|| - \bdy \big)}{\bdy} \bigg),\\
    \end{aligned}
\end{equation}
\textit{where $\erfc$ is the complementary error function defined as}
\begin{equation}
    \begin{aligned}
        \text{erfc}(z) = \frac{2}{\sqrt{\pi}} \int_z^{\infty} e^{-t^2} dt;
    \end{aligned}
\end{equation}
\textit{$\bdy$ denotes the average distance between $\cy$ and the features of the training data labeled by $y$.}

As depicted in the left of Figure \ref{fig:sensitivity_measurement}, the black periphery of the circle is the sensitive region determined by $\bdy$, and the red dot indicates the cluster center $\cy$;
the black dots represent the queries within the sensitive region, whereas the green dots are those out of the region.
Intuitively, if the query's feature $f^E(\x)$ dynamically gets close to $\cy$, $P(\x, \y)$ will then increase and be close to $1$. Inversely, if $f^E(\x)$ gets distant from $\cy$, $P(\x, \y)$ decreases and eventually end up with $0$.
Further, the SQS is going to affect the CQS computation by adjusting the radius of the queried space for each query.



\header{Cumulative Query Sensitivity.}
Given an arbitrary query sequence sharing the same predicted label, the defender aims to quantify the impact caused by honestly returning the confidence vector. 
As depicted in the right of Figure \ref{fig:sensitivity_measurement}, for all data samples in the training dataset labeled by $\y$, the defender uses $f^E$ to extract the training features. 
The training features together form a hypersphere dyed gray in the feature space $\mcl F$, which is called the \textit{sensitive space}. 
The defender then extracts the query features from the queries and treats each query feature as a small hypersphere dyed red that represents itself and the queries whose features fall around the center of the hypersphere. 
Together, the query features forms the \textit{queried space}.

By summing up the volume of the query features and dividing it by the volume of the hypersphere formed by the training data, the defender derives the ratio of the queried space to the sensitivity space. We thus define this ratio as the CQS.

\textit{Definition 3--Cumulative Query Sensitivity:} For a query sequence $X = \{\x_1,...,\x_m\}$ labeled by $y=\arg\max\y$, the cumulative sensitivity $cqs(X, y)$ is computed by
\begin{equation}\label{eq:cqs}
    \begin{aligned}
        cqs(X, y) = \frac{\sum_{\x \in X} v(sqs(\x, y), r)}{v(sqs(\cy, y), \bdy)},
    \end{aligned}
\end{equation}
where $v(\cdot, \cdot)$ is the function that computes the volume of a hypersphere; $r$ is a hyperparameter that defines the radius of the hyperspheres of the query features.

If $cqs(\x, \y)$ exceeds a pre-defined threshold $t$, this training class is considered to be threatened by the query sequence. The counterattack is launched to protect the protectee model.

\header{Output Perturbation.}
If a query sequence categorized into one class is determined to be threatening that class, the defender stops offering the true confidence vector.
Instead, the defender proactively jeopardizes the potential possibilities where the queries can support the training of a piracy model. 
The defender starts to offer falsified confidence vectors to the adversary. 
Theoretically, the falsified confidence vector is designed to make the piracy model to generate reverse gradient such that the piracy model gets worsened when the falsified confidence vector is used in the training process.
We define this objective as follows.

\textit{Definition 4--Gradient Reverse: Given an arbitrary query $\x$, the defender aims to create a falsified confidence vector $\fy$ such that}
\begin{equation}\label{eq:gradient_reverse_def}
    \begin{aligned}
        \nabla_\tau \mcl L\bigg( \sm(h(\mbf x; \tau)), \fy \bigg) 
        & = \\
        - \nabla_\tau \mcl L\bigg( \sm(h(\mbf x; \tau)), & \ \sm(f(\mbf x; \theta)) \bigg),
    \end{aligned}
\end{equation}
\textit{making the gradient completely reverse to the correct direction.}

Thereby, the piracy model trained using $(\x, \fy)$ is crippled, and therefore its performance will be far from comparable to the protectee model.





\section{QUEEN Method}

\subsection{Overview of QUEEN}\label{sect:design_of_queen}
\begin{figure*}[t!]
    \centering
    \includegraphics[width=.75\textwidth]{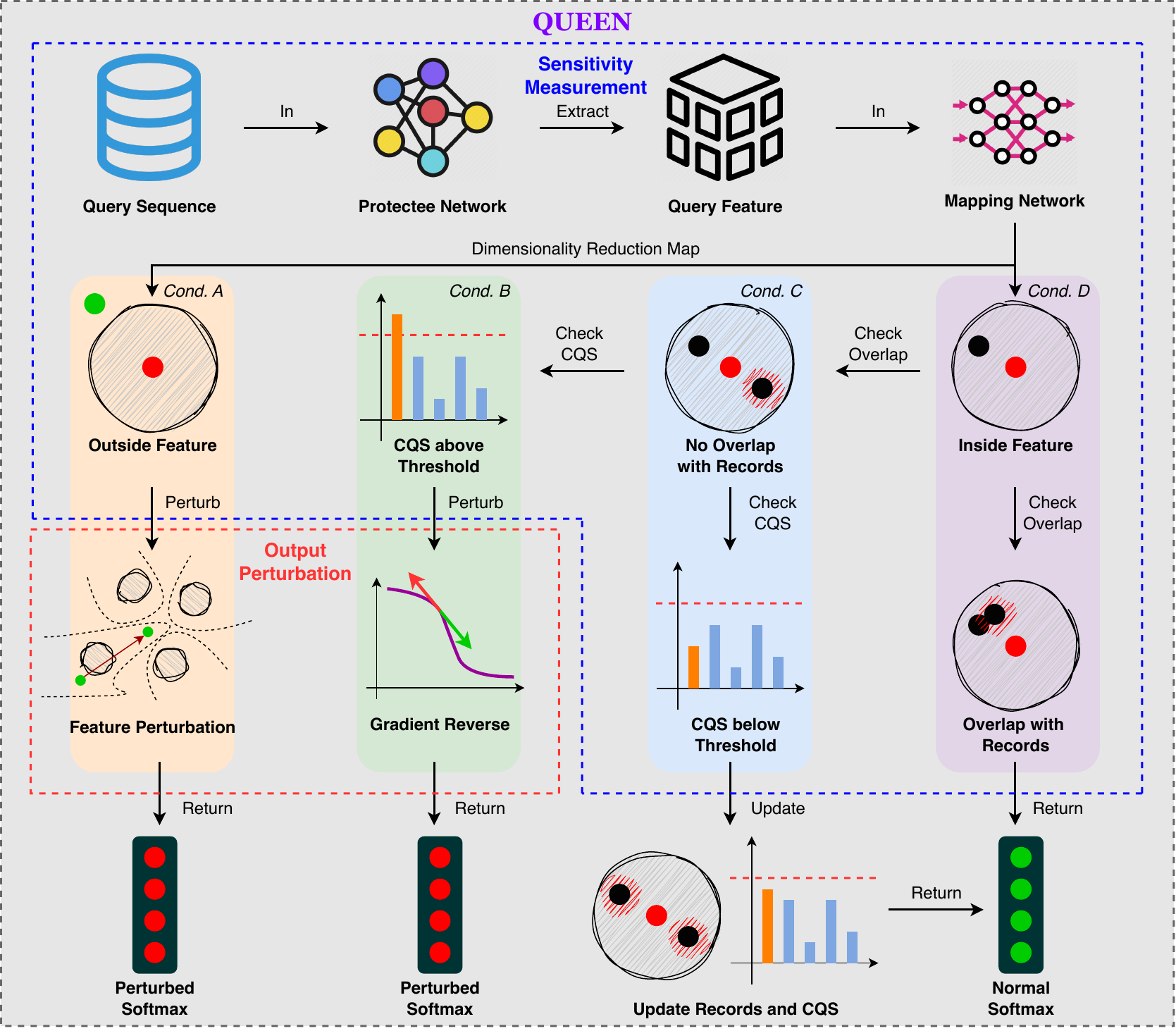}
    \caption{The workflow of QUEEN. After sensitivity measurement, there are four conditions where the output is either honestly returned or perturbed with gradient reverse or feature perturbation.}
    \label{fig:queen_workflow}
\end{figure*}

There are two main components in our method, namely \textbf{sensitivity measurement} and \textbf{output perturbation}. 
As depicted in Figure \ref{fig:queen_workflow}, given the query sequence sent by the user, the defender extracts the query feature with the protectee network.
The query feature is then further mapped to a lower-dimensional space for sensitivity measurement.
Based on the measurement, different strategies are employed to perturb the softmax output or to return the normal softmax output.
The components are further specified as follows.

In sensitivity measurement, a query sensitivity estimation system that estimates the single query sensitivity (SQS) for each query is established. The cumulative query sensitivity (CQS) can thus be derived from the SQS of each query in a given query sequence. 

In gradient reverse, given a user and a query sequence, the defender compute the CQS of the query sequence. Then, based on the CQS, the defender determines whether the query sequence is threatening the training dataset class-wise. If there is any potential that the queries of a class could support the training of a piracy model, the defender launches a counterattack by sending falsified confidence vectors. The falsified confidence vectors are designed to worsen the piracy model if they are used in the training process.

\subsection{Sensitivity Analysis}
Before making the pre-trained protectee model publicly available, the defender first performs a sensitivity analysis on every class of the training data, so that the defender is able to measure the CQS given any query sequence from the user. The process of sensitivity analysis is demonstrated in Alg. \ref{alg:sensitivity_analysis}, where the objective is to obtain a trained mapping network $g$, the set of cluster centers $\mbf c$ and the set of average distance $\bd$.

\begin{algorithm}[t]
\caption{Sensitivity Analysis}\label{alg:sensitivity_analysis}
    \KwData{$f$: the pre-trained protectee model; $g$: the mapping network; $D$: the training dataset; }
    \KwResult{$g$: the trained mapping network; $\mbf c$: the set of cluster centers; $\bd$: The set of average distances;}
    
    Use $f$ to extract the training features from $D$\;
    Train $g$ to map the training features to $\mbb R^2$ with the supervised contrastive loss till converged\;

    \For{each class in $D$}{
        Compute the cluster center\;
        Compute the average distance between the cluster center and the 2D features in this class\;
        Store the cluster center and the average distance in $\cb$ and $\bd$\;
    }
    \Return $g$, $\mbf c$, $\bd$
\end{algorithm}

The defender owns a protectee model $f$ that is trained on the training dataset $D$. The defender uses the feature extraction layers $f^E$ of $f$ to extract the training features of the training data, which is denoted as $U = \{\big((u_1, \y_1 \big),...,\big((u_n, \y_n \big)\} \subseteq \mcl F$, where $u_i = f^E(\x_i)$. However, it is not practical to compute the volume of the hypersphere in $\mcl F$, because volume computation in high-dimensional space is complex. Additionally, the defender does not want to modify the parameters $\theta$ in $f$ to realize dimension reduction. Therefore, the defender needs a fixed mapping network $g$ that maps the features from $\mcl F$ to a low-dimensional space. 

In our setting, we define $g$ as a mapping network that maps the features to a two-dimensional (2D) space, denoted a $g:\mcl F \mapsto \mbb R^2$. $g$ can have a very simple architecture, such as several fully-connected layers. Moreover, $g$ is required to make the mapped features preserve the property as that of the features in $\mcl F$. That is to say, central features should be more representative than the peripheral features. To achieve this goal, we train $g$ using the supervised contrastive loss\cite{khosla2020supervised} that takes the following form.
\begin{equation}\label{eq:super_con_loss}
    \begin{aligned}
        L^{\text{sup}} = \sum_{i \in I} -\log \bigg \{ \frac{1}{|O(i)|} \sum_{o \in O(i)} \frac{\exp(\z_i \cdot \z_o / \gamma)}{\sum_{a \in A(i)} \exp(\z_i \cdot \z_a / \gamma)} \bigg \}
    \end{aligned}
\end{equation}
Here, within a multi-view batch $U'$ randomly sampled from $U$,  $i \in I \equiv \{1,...,N\}$ is the index of samples in the batch; 
$\z$ denotes the mapped 2D feature, namely $\z = g(f^E(\x))$; The operator $\cdot$ denotes the inner product; 
$A(i) \equiv I \backslash \{i\}$ denotes the set of indices without $i$; $O(i) \equiv \{o \in A(i): \y_o = \y_i\}$ is the set of indices of all other features sharing the same label as that of $\z_i$; 
the operator $|\cdot|$ returns the number of elements in the set; $\gamma \in \mbb R^+$ is a scalar temperature parameter.
In short, by minimizing $L^{\text{sup}}$, the defender makes $g$ maps the features in the same class closely, whereas those of different classes are mapped distantly. 
Thus, the relative position of feature $u_i \in U$ is kept when it is mapped to $\z_i$.
Another feasible solution is to use the hierarchical contrastive loss proposed in \cite{zhang2022HierarchicalContrastiveLoss}.

With a trained mapping network $g$, the defender is able to map entire training feature set $F$ to a 2D feature set denoted by $Z = \{(\z_1, \y_1),...,(\z_n, \y_n)\}$. 
Let $Z^y \subseteq Z$ denote the 2D training feature set of class $y = \arg\max (\y)$. 
The defender now computes the cluster center $c^y$ in the 2D space for each $Z^y$ as defined in Eq. \ref{eq:cluster_center} by
\begin{equation}
    \begin{aligned}
        c^y = \frac{1}{|Z^y|} \sum_{\z_i \in Z^y} \z_i,
    \end{aligned}
\end{equation}
Meanwhile, the defender also computes the average distance $\bdy$ between $c^y$ and $\z_i \in Z^y$ by
\begin{equation}
    \begin{aligned}
        \bdy = \frac{1}{|Z^y|} \sum_{\z_i \in Z^y} \|c^y - \z_i\|.
    \end{aligned}
\end{equation}

\subsection{Sensitivity Measurement}\label{sct:sensitivity_measurement}
\begin{algorithm}[t]
\caption{Sensitivity Measurement}\label{alg:sensitivity_measurement}
    \KwData{$f$: the pre-trained protectee model; $D'$: the query sequence;}
    \KwResult{$\s$: the set of CQS of each class;}

    Use $f$ to get predicted labels for queries in $D'$\;
    Split $D'$ by the predicted labels\;
    \For{each split query sequence}{
        Determine the sensitive region by $\bdy$\;
        \For{each query}{
            Get the 2D feature of the query\;
            \If{the feature is within the sensitive region}{   
                \If{no overlapping records}{
                    Compute its SQS\; 
                    Record this feature\;
                }
            }
        }
        Get the queried area by summing up the area of each sensitive feature circle\;
        Get the CQS of this class by dividing the queried area by the area of the sensitive region\;
        Store the CQS in $\s$\;
    }
    \Return $\s$
\end{algorithm}
With the mapping network $g$, the set of cluster center $c = \{...,c^y,...\}$ and the set of average distances $\bd = \{...,\bdy,...\}$, the defender can measure the sensitivity of any given query sequence. 
As illustrated in Alg. \ref{alg:sensitivity_measurement}, given the auxiliary dataset $D' = \{\x_1,...,\x_m\}$, the defender gets the set of CQS $\s = \{...,\s^y,...\}$, where $\s^y$ denotes the CQS of class $y$.

The defender first gets the predicted label $\yh_i = \arg \max \sm(f(\x_i')), \forall \ \x_i' \in D'$. 
Next, the defender splits $D'$ by the predicted labels such that $D'^{\yh}$ denotes the subset of $D'$ labeled by $\yh$. 
The sensitivity measurement is then conducted class-wise. 
For each class, the defender determines the sensitive region by setting its radius to $\bd^{\yh}$.

To compute the SQS, the defender gets the 2D feature $\z_i = g(f^E(\x_i')), \forall \ \x_i' \in D'^{\yh}$. 
The defender then needs to determine whether $\z_i$ satisfies the following two conditions: 
\textit{1. $\z_i$ is within the sensitive region, i.e., $\|\z_i - c^{\yh}\| < \bdyh$; 
2. There is no previously recorded feature overlapping with $\z_i$, i.e., $\nexists z_j \in Z^{\yh} \text{, s.t. } \|z_i - z_j\| < r, i \neq j$.} 
Here, 
$r$ is a hyperparameter denoting the radius of each query circle representing the features of the query and those nearby it. If the conditions are satisfied, the defender proceeds to compute the SQS of the query by
\begin{equation}
    \begin{aligned}
        \sqs(\z_i, \yh) = \frac{1}{2} \erfc \bigg( \frac{\|\z_i - c^{\yh}\| - \bd^{\yh}}{\bd^{\yh}} \bigg).
    \end{aligned}
\end{equation}

Eventually, the CQS of the 2D features satisfying the conditions $Z$ with the predicted label $\yh$ is computed by
\begin{equation}
    \begin{aligned}
        \cqs(Z, \yh) & = \frac{r^2}{(\bdyh)^2} \sum_{\z_i \in Z^{\yh}} (\sqs(\z_i, \yh))^2.
    \end{aligned}
\end{equation}
Intuitively, $\cqs(Z, \yh)$ indicates the ratio of the queried area to the sensitive area in class $\yh$. If $\cqs(Z, \yh)$ exceeds a pre-defined threshold, the adversary is determined to be threatening class $\yh$. Thus, the defender ought to launch a counterattack to prevent further losses.

Eventually, there are four conditions after the sensitivity measurement as depicted in Figure \ref{fig:queen_workflow}:
\begin{itemize}
    \item \textit{Cond. A: the 2D feature is not within the sensitive region.}

    \item \textit{Cond. B: the 2D feature is within the sensitive region; the 2D feature overlaps with the records; the CQS exceeds the threshold.}

    \item \textit{Cond. C: the 2D feature is within the sensitive region; the 2D feature does not overlap with the records; the CQS is below the threshold.}

    \item \textit{Cond. D: the 2D feature is within the sensitive region; the 2D feature overlaps with the records.}
\end{itemize}
In \textit{Cond. A/B}, the defender launches feature perturbation/gradient reverse to perturb the softmax outputs, whereas the normal softmax outputs are returned in \textit{Cond. C/D}. 

We claim that the data points whose features are not in the sensitive region can also be used in training the piracy model.
Our CQS measurement does not include those data points.
But those data points are also perturbed by feature perturbation to reduce their contribution to the piracy model training.

\subsection{Output Perturbation}
As demonstrated in Alg. \ref{alg:output_perturbation}, the defender proactively defends any potential attacks by conditionally perturbing the confidence vector generated by the protectee model.
Given a pre-defined threshold $t$, if $\cqs(Z, \yh) > t$, the defender starts the gradient reverse counterattack by sending a falsified confidence vector to the adversary to achieve the objective described in Eq. \ref{eq:gradient_reverse}. To achieve this objective, the defender needs two main components in gradient reverse, namely \textit{piracy model simulation} and \textit{gradient reverse}. 

For the query whose feature is not within the sensitive region, the defender performs \textit{feature perturbation} by moving the feature towards the most distant cluster center. The perturbation stops before the predicted label of the perturbed feature changes, resulting in a confidence vector that worsens the piracy model trains with it.

\begin{algorithm}[t]
\caption{Output Perturbation}\label{alg:output_perturbation}
    \KwData{
        $f$: the protectee model;
        $D$: the training dataset;
        $D'$: the query sequence;
        $t$: the threshold;
        }
    \KwResult{$Y$: The confidence vectors;}

    \Comment{Piracy Model Simulation}
    Randomly split $D$ into subsets evenly\;
    Train a set of shadow models of different architectures, and each shadow model is only trained on one subset without intersection\;

    \Comment{Logits Falsification}
    \For{$\x \in D'$}{
        Extract the 2D feature of $\x$ using $f$ and $g$\;
        Get the softmax $\ybh \gets \sm(f(\x))$\;
        Get the label $\yh \gets \arg \max \ybh$\;

        Measure the sensitivity of $\x$\;
        \eIf{sensitive}{
            Perform gradient reverse or feature perturbation based on the conditions to get the falsified softmax output $\ybt$\;
            Store $\ybt$ in Y\;
        }
        {
            Store $\ybh$ in Y\;
            Update feature records and CQS accordingly\;
        }
        

        
    }
    
    \Return $Y$
    
\end{algorithm}

\header{Piracy Model Simulation.} Due to the fact that the defender has zero knowledge about the piracy model, the defender cannot launch gradient reverse because the output of the piracy model is not accessible. To address this problem, the defender needs piracy model simulation, i.e., to train a set of shadow models locally to simulate the behavior of the piracy model. The weights of different models typically converge to the initial stable points within the same optimization problem \cite{Haim2022ReconstructingTrainingData}, indicating that the gradient descent directions of the shadow and piracy models are similar. Similarly, the use of shadow models to mimic the piracy model is also illustrated in \cite{orekondy2019prediction}. 

The training dataset $D$ is randomly divided into a number of subsets that are used to train shadow models. The shadow models are of various architectures, and each shadow model is paired with one subset without overlapping. In order to simulate the piracy model with randomness, a fixed number of the shadow models are drawn and the average of their logits is used to estimate the piracy model's logit given this input.

\header{Gradient Reverse.}
Given an arbitrary query $\x \in D'$, the defender first extracts the 2D feature $\z = g(f^E(\x))$, and gets the softmax $\ybh = \sm(f(\x))$ and the predicted label $\yh = \arg \max \ybh$. 
If $\x$ is within the sensitive region, and there is no previous record overlapping with it, the query is sensitive. 
Next, the defender checks whether the CQS of class $\yh$ exceeds the pre-defined threshold $t$. 
If yes, the defender starts logits falsification, otherwise it will be treated as a normal query, and $\ybh$ will be honestly returned. 
The feature of the sensitive query is recorded for overlapping checking.

To perturb the softmax for the sensitive query, the defender randomly draws a subset $H'$ of shadow models from the trained shadow model set $H$. The falsified logits $\ybh'$ is derived by
\begin{equation}
    \begin{aligned}
        \ybh' = \frac{1}{|H'|} \sum_{h \in H'} \sm(h(\x)),
    \end{aligned}
\end{equation}
which is the mean of the logits from the shadow models. With $\ybh$ and $\ybh'$, the defender can now create a falsified logit $\ybt$ that fulfills the objective of gradient reverse by
\begin{equation}\label{eq:gradient_reverse}
    \begin{aligned}
        \ybt = 2 \ybh' - \ybh.
    \end{aligned}
\end{equation}

In practice, we need to solve an optimization problem so as to get a valid $\ybt$, because $\ybt$ derived through Eq. \ref{eq:gradient_reverse} is often not a valid softmax output that can be easily detected by the adversary.
The optimization problem is formulated as follows:
\begin{equation}
    \begin{aligned}
        \underset{\ybt}{\max} \ & \text{cossim}(\ybt, 2\ybh' - \ybh),\\
        \text{subject to } & \sum_{\tilde{y} \in \ybt} \tilde{y} = 1,\\
        & \tilde{y} \geq 0, \ \forall \tilde{y} \in \ybt.
    \end{aligned}
\end{equation}
After the optimization, $\ybt$ becomes a valid softmax output where each element in it is greater than $0$, and the sum of them is equal to $1$.
The direction of $\ybt$ is also optimized to be as close as possible to that of $2\ybh' - \ybh$.

\begin{algorithm}[t]
\caption{Feature Perturbation}\label{alg:feature_perturbation}
    \KwData{
        $\x$: the query;
        $\mbf c$: the cluster centers;
        $f$: the protectee model;
        }
    \KwResult{$\ybt$: The perturbed output;}

    $u \gets f^E(\x)$\;
    $\yh \gets \arg \max \sm(f^C(u))$\;
    $y \gets \underset{y}{\arg \max} \| \z - \cy \|$\;
    $v \gets \frac{\z - \cy}{\| \z - \cy \|}$\;
    \While{True}{
        $u' \gets u + \epsilon v$\;
        \eIf{$\arg \max \sm(f^C(u')) == \yh$}{
            $u \gets u'$\;
        }{
            Break\;
        }
    }
    $\ybt \gets \sm(f^C(u))$\;
    
    \Return $\ybt$
    
\end{algorithm}

\header{Feature Perturbation.}
Given a query $\x$ with the predicted label $\yh = \arg \max f(\x; \theta)$ that is not in the sensitive region, i.e., $\| g(f^E(\x)) - c^{\yh} \| > \ryh$, the defender creates a falsified confidence vector $\ybt$ by iteratively perturbing the query feature.

Let $u = f^E(\x)$ denote the query feature in $\mcl F$. The defender first finds out the most distant cluster center in $\mbf c$ to $u$ by computing
\begin{equation}
\begin{aligned}
    \underset{y}{\arg \max} \ \| u - \cy \|,
\end{aligned}
\end{equation}
where $\cy \in \mcl F$ is computed via Eq. \ref{eq:cluster_center}.
Next, the defender perturbs $u$ by moving $u$ towards $\cy$ stepwise until the predicted label of $u$ changes. Let $v = \frac{u - \cy}{\| u - \cy \|}$ be the vector of movement, the perturbation process is defined as
\begin{equation}
\begin{aligned}
    u \gets u + \epsilon v, \text{ s.t. } \arg \max \sm(f^C(u)) = \yh
\end{aligned}
\end{equation}
where $\epsilon$ denotes the step size of the perturbation.
Eventually, the defender returns $\ybt = f^C(u)$ to the user. Intuitively, feature perturbation aims to obfuscate the decision boundary of the piracy model by making $f(\x; \theta)$ appear at the edge of the two most irrelevant classes, thereby damaging the decision boundary in a similar way compared to boundary unlearning \cite{chen2023boundaryunlearning}.

\section{Theoretical Analysis}

The proofs of the following theorems are given in Section \ref{sect:supp_theo_als} in the Supplemental Materials.

\subsection{Feasibility of Gradient Reverse}\label{sect:proof_gradient_reverse}

\header{Theorem 1.}
\textit{(The sufficient condition of gradient reverse) To achieve gradient reverse in Definition 4, it is enough to set  $\ybt = 2\ybh' - \ybh$.}



\subsection{Certifiability}

The aim of the defender is to control the number of honestly answered queries $\eta$.
Thus, the essential problem is to estimate the maximum number of honestly answered queries $\hat{\eta}$, given the maximum allowable error $\epsilon$ with the error probability $\delta$ by the defender. 
In other words, \textbf{if the defender pre-defines $\epsilon$ and $\delta$, the defender knows how to set the threshold $t$ and radius $r$}, because $\eta$ can be estimated by $t$ and $r$.
To achieve this aim, we would like to theoretically identify the relationship between $\epsilon$, $\delta$, $t$ and $r$ with the help of the probably approximately correct (PAC) learning theory~\cite{valiant1984theory}.



\header{Settings.}
Suppose the adversary uses its auxiliary dataset $D'$ containing $\eta$ samples to query the protectee model $f$ and collects the related softmax outputs as ground-truth labels. 
Within the sensitive region, all softmax outputs are honestly given by $f$.
Let $h$ denote the piracy model trained on $D'$ with respect to the query results at its finest.
We consider the true error $E(h)$ as the probability that $h$ makes a mistake on the sample $(\x, \y)$ from the problem domain $\mcl D$, $E(h)=\Pr_{(\x, \y) \sim \mathcal D}[h(x) \ne y]$, and similarly $E(h_e)$ is the empirical error that describes the mistake made by $h$, $E(h_e)=\frac{1}{\eta}\sum_{x \in D'}\mathbf 1[h(\x) \ne f(\x)]$.
Let $t$ denote the threshold and $r$ refers to the radius.
Intuitively, the defender can estimate $t$ and $r$ if he or she has an expectation on how much the adversary can learn from the queries.

\header{Theorem 2.}
\textit{Given a learning algorithm that learns a piracy model $h$, let $\eta$ be the actual number of honestly answered sensitive queries, the piracy model can at most be trained to have the maximum allowable error $\epsilon$ and the upperbound of error probability $\delta$, if}
\begin{equation}
\begin{aligned}
    \eta \leq \hat{\eta} = \frac{1}{2\epsilon^2}\cdot\ln(\frac{2}{\delta}),
\end{aligned}
\end{equation}
\textit{where $\hat{\eta}$ is the maximum number of honestly answered sensitive queries.}
\textit{Further, let $t$ be the threshold, and $r$ be the query radius, we will have the relationship as follows.}
\begin{equation}
\begin{aligned}
    r \geq \sqrt{\frac{2t}{\ln(\frac{2}{\delta})}}\epsilon\bd,
\end{aligned}
\end{equation}
\textit{where $\bar d$ is a constant denoting the average distance of features to their cluster center.}





Furthermore, we provide another proof that justifies the effectiveness of QUEEN.

\header{Theorem 3.} \cite{tangmodelguard,cover1999elementsOfInfoTheory}
\textit{
Any adaptive model extraction attack with an arbitrary recovery function $R(\cdot)$ cannot attain a smaller gap between the recovered predictions $R(\tilde{Y}) \in \mbb R^{M\times N}$ and the original predictions $Y \in \mbb R^{M \times N}$ than the following lower bound:
}
\begin{equation}
\begin{aligned}
    \mbb E[\|R(\tilde{Y}) - Y\|^2_2] \geq \frac{MN}{2\pi e} \exp{\bigg(\frac{2}{MN} h(Y|\tilde{Y}) \bigg)},
\end{aligned}
\end{equation}
\textit{
where $M$ and $N$ respectively denote the number of the samples and classes;
$h(Y|\tilde{Y})$ is the conditional entropy.
}





\section{Experiment}
\subsection{Experiment Settings}
The details of datasets, model architectures, and the implementation of QUEEN are in Section \ref{sect:exp_settings} in the Supplemental Materials.
We introduce the attacks and defenses used in the experiment as follows.

\header{Attacks.}
A model extraction attack consists of two parts:
the query strategy and attack strategy.
In other words, different attack strategies can be combined with various query strategies.

Two query strategies are considered in this experiment: JBDA-TR \cite{juuti2019prada} and KnockoffNet \cite{orekondy2019knockoff}.
KnockoffNet uses only natural data in the auxiliary dataset, whereas JBDA-TR synthesize data from a small seed set sampled from the auxiliary dataset using Jacobian-based data augmentation \cite{papernot2017practical}.
We set the query budget of KnockoffNet to $50,000$ and $1,000$ for the size of the seed set of JBDA-TR to ensure that it suffices to allow the adversary to get the best attack accuracy.
We utilize the random knockoff-net configuration as described in \cite{orekondy2019knockoff,tangmodelguard}.

The employed attack strategies are listed as follows: 
\begin{enumerate}
    \item \textbf{Direct Query}: The adversary directly uses the output from the protectee model to train the piracy model.

    \item \textbf{Label-Only}: The top-1 hard label is kept, whereas the other results in the softmax output are ignored in the training process.

    \item \textbf{S4L} \cite{jagielski2020high}: Each query image is differently augmented multiple times such that the query results of the different versions of the original query image are averaged to recover the perturbed outputs.

    \item \textbf{Smoothing} \cite{lukas2022sok}: Each query image is augmented by random affine augmentations. Similar to S4L attack, the query results of different augmented versions of the same image are eventually averaged to recover the perturbed outputs.

    \item \textbf{D-DAE} \cite{chen2023ddae}: Meta-classifiers are used to identify the defense method employed by the defender.
    A number of shadow models are trained to generate clean outputs.
    A generative model is then trained to recover the perturbed output to the normal output.

    \item \textbf{D-DAE+} \cite{tangmodelguard}: This is the improved version of D-DAE, because it utilizes the lookup table generated for Partial Bayes Attack (listed below) as the training data of the generative model.

    \item \textbf{Partial Bayes (pBayes)} \cite{tangmodelguard}: This attack employs independent sampling and the neighborhood sampling to establish the Bayes estimator to recover the perturbed outputs.
\end{enumerate}

\header{Defenses.}
The following defenses are evaluated in our experiments:
\begin{enumerate}
    \item \textbf{None}: No defense method is employed.
    Clean softmax outputs are directly returned to the adversary.

    \item \textbf{Reverse Sigmoid (RS)} \cite{lee2018defending}: The clean outputs are perturbed by a reverse sigmoid function to enlarge the cross entropy loss.

    \item \textbf{Maximizing Angular Deviation (MAD)} \cite{orekondy2019prediction}: The clean outputs are perturbed by maximizing the angular deviation between the gradients calculated using the clean outputs and the perturbed outputs.

    \item \textbf{Adaptive Misinformation (AM)} \cite{kariyappa2020defending}: The protectee \newline model is trained to generate perturbed outputs when it meets Out-Of-Distribution (OOD) queries.

    \item \textbf{Label-Only}: The hard top-1 label is sent back to the user.

    \item \textbf{Rounding}: The confidence score in the softmax output is pruned to have only one decimal place. For example, $[0.14, 0.46, 0.31, 0.09] \gets [0.1, 0.5, 0.3, 0.1]$.

    \item \textbf{Exponential Mechanism Differential Privacy (EMDP)} \cite{Ilvento2020ExponentialMechanism}: The softmax output is perturbed by adding noises generated via exponential mechanism based differential privacy.

    \item \textbf{ModelGuard} \cite{tangmodelguard}: The ModelGuard here refers to \newline ModelGuard-W in the original paper, because ModelGuard-W is better than ModelGuard-S in most cases. The clean output is perturbed by solving a constrained optimization problem that tries to maximize the cross-entropy loss when the adversary trains the piracy model with the perturbed outputs.

    \item \textbf{QUEEN}: The query's sensitivity is measured and the output is perturbed as described in Section \ref{sect:design_of_queen}.
\end{enumerate}




\begin{table*}[t!]
    \caption{Evaluation of defenses against attacks on CIFAR10.}
    \label{tab:eval_cifar10}
    \centering
    \begin{tabular}{ccccccccccc}
        \toprule
        Query Method & Attack Method & None & RS & MAD & AM & Label-only & Rounding & EMDP & ModelGuard & QUEEN \\
        \midrule
        \multirow{7}{*}{KnockoffNet} & Direct Query & 87.42\% & 85.33\% & 84.58\% & 83.17\% & 83.78\% & 86.77\% & 66.15\% & 74.88\% & \textbf{10.00\%} \\
         & Label-Only & 83.78\% & 83.78\% & 83.78\% & 82.11\% & 83.78\% & 83.78\% & 83.78\% & 83.78\% & \textbf{81.17\%} \\
         & S4L & 86.17\% & 82.30\% & 80.21\% & 82.12\% & 84.02\% & 85.86\% & 66.76\% & 70.69\% & \textbf{10.00\%} \\
         & Smoothing & 65.43\% & 63.41\% & 61.23\% & 62.27\% & 61.01\% & 65.10\% & 64.36\% & 53.24\% & \textbf{10.00\%} \\
         & D-DAE & 87.42\% & 85.32\% & 84.36\% & 78.38\% & 85.24\% & 87.45\% & 71.43\% & \textbf{64.73\%} & 78.21\% \\
         & D-DAE+ & 87.42\% & 85.91\% & 86.44\% & 84.51\% & 84.55\% & 87.01\% & 86.43\% & 58.17\% & \textbf{50.24\%} \\
         & pBayes & 87.42\% & 85.91\% & 87.24\% & 86.93\% & 84.57\% & 86.99\% & 85.41\% & 85.16\% & \textbf{84.24\%} \\
         \hline
         \multirow{5}{*}{JBDA-TR} & Direct Query & 63.51\% & 67.01\% & 55.31\% & 60.86\% & 63.31\% & 73.55\% & 25.92\% & 37.91\% & \textbf{10.00\%} \\
         & Label-Only & 63.51\% & 63.51\% & 63.51\% & \textbf{55.77\%} & 63.51\% & 63.51\% & 63.51\% & 63.51\% & 61.45\% \\
         & D-DAE & 74.41\% & 56.63\% & 48.51\% & 57.17\% & 60.15\% & 67.65\% & 62.17\% & 59.17\% & \textbf{47.10\%} \\
         & D-DAE+ & 74.41\% & 72.48\% & 68.33\% & 66.10\% & 63.44\% & 73.07\% & 62.33\% & 51.85\% & \textbf{40.88\%} \\
         & pBayes & 74.41\% & 71.21\% & 68.15\% & 74.11\% & 65.42\% & 75.01\% & 67.93\% & 65.54\% & \textbf{65.33\%} \\
         \hline
         \multicolumn{2}{c}{Max Piracy Model Accuracy} & 87.42\% & 85.91\% & 87.24\% & 86.93\% & 85.24\% & 87.45\% & 86.43\% & 85.16\% & \textbf{84.24\%} \\
         \multicolumn{2}{c}{Max Piracy Model Agreement} & 88.24\% & 87.21\% & 89.01\% & 88.22\% & 87.13\% & 88.67\% & 88.71\% & 86.78\% & \textbf{86.54\%} \\
         \multicolumn{2}{c}{Protectee Model Accuracy} & 92.74\% & 92.74\% & 92.74\% & 90.15\% & 92.74\% & 92.74\% & 92.74\% & 92.74\% & 90.01\% \\
         \bottomrule
    \end{tabular}
\end{table*}

\subsection{Experimental Results}
In this section, two groups of experiments are presented.
First, we evaluate the effectiveness of QUEEN along with the other defenses against various attacks with fixed hyperparameters.
It is observed that QUEEN outperforms the other defenses in most cases in the same utility scope.
Second, we test how the hyperparameters of QUEEN trade off the performance of the protectee model and the piracy model.


\header{Effectiveness of QUEEN.}
The defenses are evaluated with the following hyperparameters.
$\epsilon$ of MAD, ModelGuard is set to be $1.0$ such that the accuracy of the protectee model is not affected, while the perturbation is maximized.
For the same reason, we set $\gamma = 0.2$ and $\beta = 0.2$ for RS.
$\tau$ of AM is set to be $0.3$ to ensure that the accuracy of the protectee model does not drop significantly.
$\epsilon$ of EMDP is set to be $1.0$.
Similarly, we set $r=0.005$ and $t=0.2$ for QUEEN to avoid a drastic decrease in accuracy.

The experimental results of effectiveness conducted on CIFAR10 are presented in Table \ref{tab:eval_cifar10}.
The results of the experiments conducted on CIFAR100, Caltech256 and CUB200 are in Table \ref{tab:eval_cifar100}, \ref{tab:eval_caltech256} and \ref{tab:eval_cub200} in the Supplemental Materials.
Each row in the table contains the accuracy of the piracy model derived by launching the attack of this row on the defense of the column.
At the bottom of the tables, we list the maximum accuracy and agreement of the piracy model among all attacks and the accuracy of the protectee model guarded by the defenses.

Based on the experimental results, we observe that QUEEN outperforms the other state-of-the-art defenses in most cases.
For the attacks that do not drastically modify the perturbed output such as Direct Query, S4L and Smoothing, QUEEN can make the piracy model have random-guessing accuracy.
Although the accuracy of the protectee model is decreased, this can be considered as the cost of the defense performance.
However, when facing the recovery-based attacks, namely D-DAE, D-DAE+, and pBayes, the accuracy of the piracy model increases significantly.
Compared to the other defenses, the perturbation scheme of QUEEN is still effective, but not strong enough to completely defend against such attacks.
Because the generative model of the recovery-based attacks can bypass the perturbation-based defenses to an extent.

Following the most recent machine unlearning evaluation metric proposed by Maini et al. \cite{maini2024tofu}, we use the forget quality to measure the effectiveness of the defenses.
We generate two groups of predicted labels by feeding the query dataset into the piracy model trained with the perturbed output and the piracy model trained with the original output, which are defined as the defended output, and the undefended output.
Different from the truth ratio used in \cite{maini2024tofu}, we 
perform Kolmogorov-Smirnov (KS) test to make a statistical test between the two histograms of the defended output and the undefended output to test whether their difference is significant.

\begin{figure*}[t]
    \centering
    \includegraphics[width=0.75\textwidth]{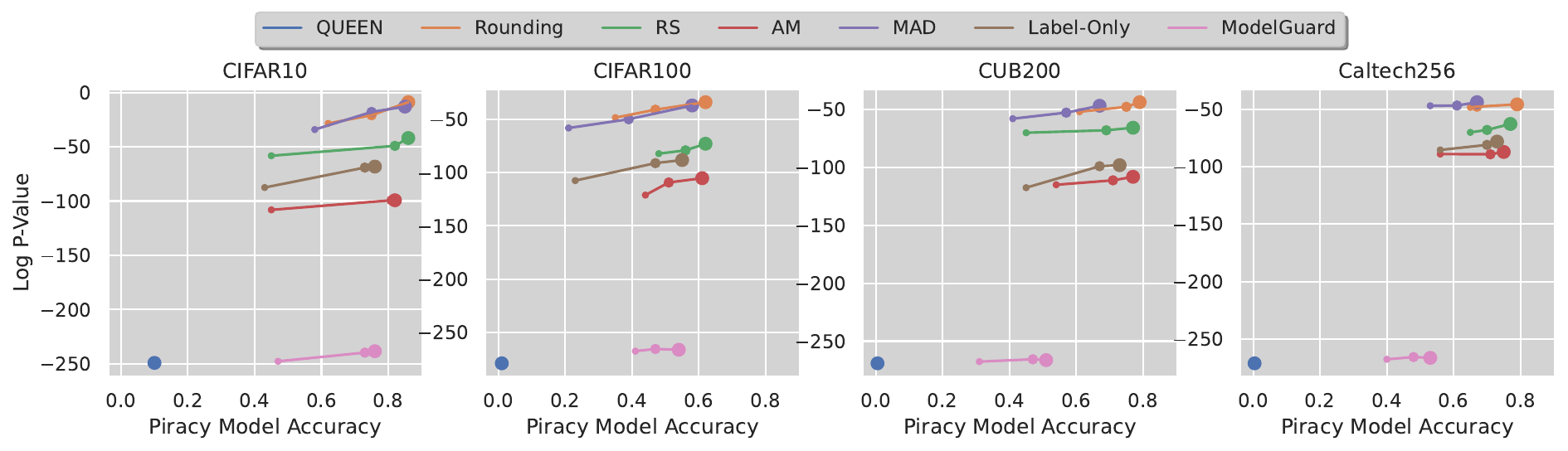}
    \caption{The results of KS test on the outputs of the piracy models.}
    \label{fig:ks_test}
\end{figure*}

As shown in Figure \ref{fig:ks_test}, the x and y axes denote the accuracy of the piracy model and the $p$-value derived from the KS test.
The size of the point represents the number of training epoch.
It is observed that QUEEN makes the piracy model have only random-guessing level performance when facing direct query attack.
The low $p$-value of Queen reflects that the difference between the output of the piracy models trained with the perturbed and the original query outputs is significant.
Notably, ModelGuard also provides a very low p-value.

\subsection{Ablation Study}\label{sect:ablation_study}
\header{Impact of the Hyperparameters.}
We further test how the \newline selection of threshold $t$ is going to affect the performance of QUEEN.
We launch KnockoffNet with D-DAE attack on the protectee model trained on CIFAR-10 using TinyImageNet200 as the auxiliary dataset.
The query budget is set to $50,000$, and $t$ varies in the range of $\{0.1, 0.2, 0.3, 0.4, 0.5\}$ with $r=0.005$.

\begin{figure*}[t!]
    \centering
    \includegraphics[width=.75\textwidth]{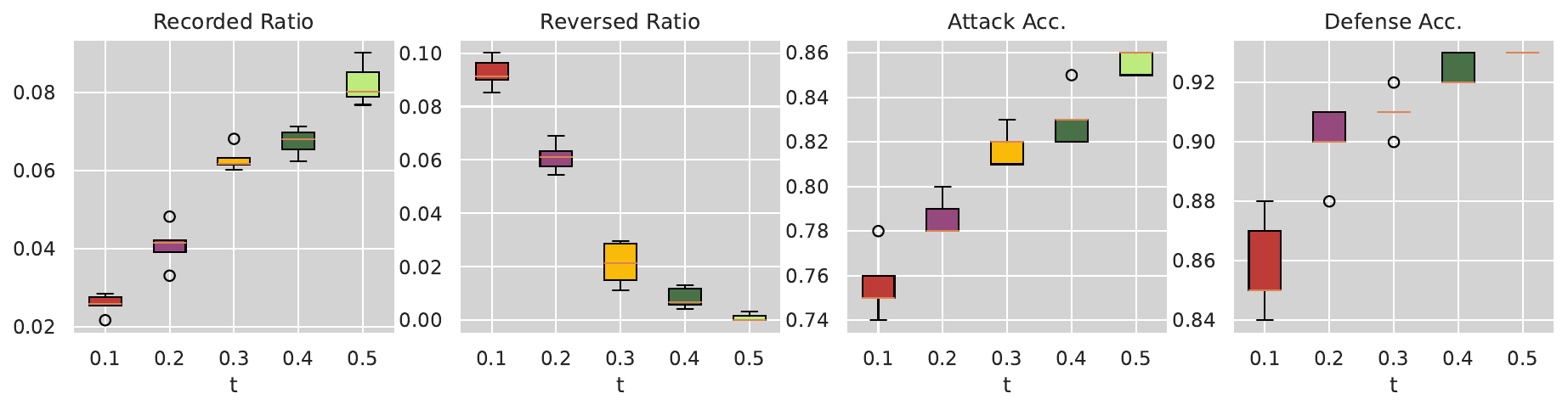}
    \caption{The impact of selection of the threshold $t$ on CIFAR-10.}
    \label{fig:ablation_threshold}
\end{figure*}

The results are shown in Figure \ref{fig:ablation_threshold}, where we test how the varying $t$ specifically affects the recorded ratio, reversed ratio, attack accuracy and defense accuracy.
For each $t$ value, we repeat the experiment $5$ times.
The recorded ratio refers to the number of recorded features over the query budget.
Similarly, the reversed ratio is the number of queries that are gradient-reversed over the query budget.
We omit the results of $t>0.5$, because a larger $t$ does not lead to any change compared to $t=0.5$.
With $t=0$, QUEEN performs gradient reverse on every query that hits the sensitive region, which leads to the theoretically lowest attack and defense accuracy.
However, this prevents the normal users from getting the honest answer from the protectee model.
Hence, we do not consider the $t=0$ situation in this experiment.
It is observed that as $t$ increases, the reversed ratio decreases, and the recorded ratio increases.
This means that more queries in the sensitive region are honestly answered, which results in both the ascending attack and defense accuracy.
In conclusion, a larger $t$ leads to a lower attack accuracy at the cost of the inevitable decrease in defense accuracy.

\begin{figure*}[t!]
    \centering
    \includegraphics[width=.75\textwidth]{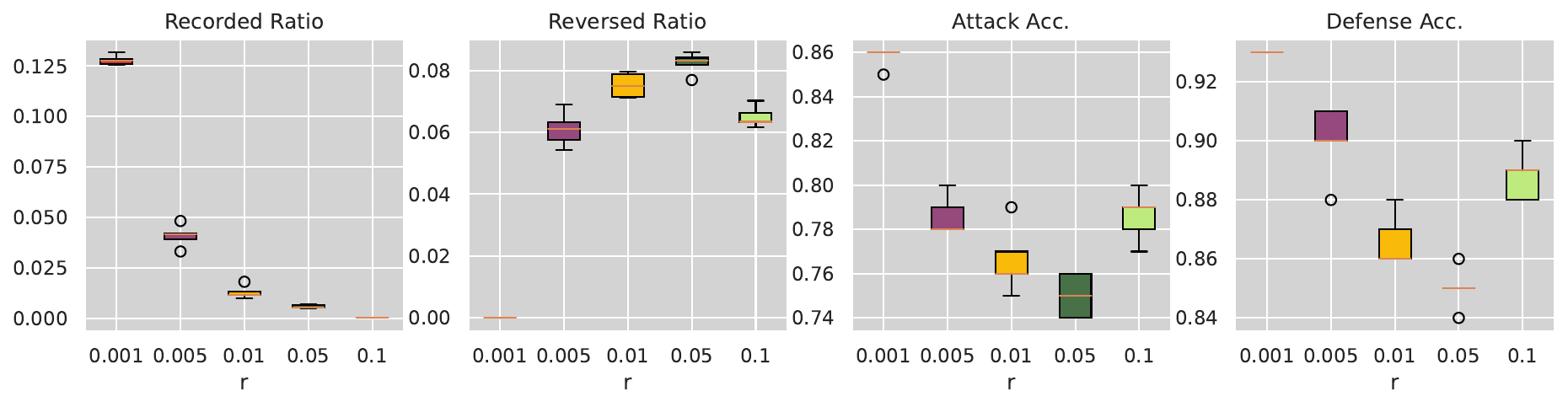}
    \caption{The impact of selection of the radius $r$ on CIFAR-10.}
    \label{fig:ablation_radius}
\end{figure*}

Next, we evaluate the impact of the selection of the radius $r$.
We set $r \in \{0.001, 0.005, 0.01, 0.05, 0.1\}$, while keeping $t=0.2$.
The rest of the experiment settings remain the same as above.
As depicted in Figure \ref{fig:ablation_radius}, we observe that as $r$ increases, the recorded ratio drops, because the area of the query increases such that the threshold is more quickly reached.
For the same reason, one recorded query thus has more neighbor queries, resulting in a decreasing reversed ratio when $r$ is too large.
This explains the inverse tendency in reversed ratio, attack accuracy, and defense accuracy when $r=0.1$.
It means that selecting an appropriate $r$ can provide a better trade-off between the performance of the defense and the model's utility.
Based on the results, we notice that when $t=0.2$ and $r=0.05$, QUEEN provides the most balanced defense, where the defense accuracy does not significantly drop, and the attack accuracy is effectively reduced.

We then investigate how different numbers of shadow models impact QUEEN by testing numbers in $\{1,2,3,5,10\}$ on the CIFAR10 dataset. The results are detailed in Table \ref{tab:num_shadow}. It is observed that the number of shadow models does not significantly affect the defense effectiveness. We also evaluate the impact of the number of shadow models on the CIFAR10 and CIFAR100 dataset using D-DAE+ and pBayes attacks, and the results are in Table \ref{tab:num_shadow_cifar10_pbayes}, \ref{tab:num_shadow_cifar100_ddaeplus} and \ref{tab:num_shadow_cifar100_pbayes} in the Supplemental Materials.

\begin{table}[t!]
    \centering
    \caption{Impact of the Number of Shadow Models on the Attack Accuracy of the Piracy Models on CIFAR10 Attacked by D-DAE+.}\label{tab:num_shadow}
    \setlength{\tabcolsep}{1.5mm}{
    \begin{tabular}{c|c|c|c|c|c}
    \toprule
    Number of Models & 1 & 2 & 3 & 5 & 10 \\
    \midrule
    Attack Accuracy & 52.09\% & 52.59\% & 52.56\% & 51.27\% & 50.24\%\\
    Attack Agreement & 52.57\% & 53.00\% & 53.13\% & 52.04\% & 50.61\%\\
    \bottomrule
    \end{tabular}}
\end{table}

\begin{table}[t!]
    \caption{Runtime Test.}
    \label{tab:runtime}
    \centering
    \scriptsize
    \renewcommand\arraystretch{1.}
    \setlength{\tabcolsep}{3mm}{
    \begin{tabular}{c|cccc}
        \toprule
        \multirow{3}{*}{Task} & \multicolumn{3}{c}{Dataset} \\
        & MNIST & CIFAR10 & CIFAR100 \\
        & \multicolumn{3}{c}{Runtime (s)} \\
        \midrule
        Training Feature Extraction & \multirow{1}{*}{12.37} & \multirow{1}{*}{15.30} & \multirow{1}{*}{16.21} \\
        
        Mapping Network Training & \multirow{1}{*}{63.89} & \multirow{1}{*}{66.91} & \multirow{1}{*}{75.47} \\
        
        Sensitivity Analysis & 1.44 & 1.39 & 1.47 \\

        Training $10$ Shadow Models & 82.79 & 94.18 & 96.41 \\

        Process 1,000 Queries & 2.51 & 4.28 & 4.61 \\
        \bottomrule
    \end{tabular}}
\end{table}

\header{Runtime.}
As demonstrated in Table \ref{tab:runtime}, we test the time consumption in terms of the preparation and evaluation of QUEEN.
The test is conducted over three datasets, and the results of each dataset are close to each other.
This is because of the number of samples in these training datasets are equal to $60,000$.
It is observed that the training feature extraction takes no longer than $17$ seconds, and training the mapping network costs less than $76$ seconds.
Additionally, the time of sensitivity analysis and query processing is trivial.
The comparison of time cost between the defenses are listed in Table \ref{tab:query_time_cifar10} in the Supplemental Materials.
Overall, the runtime is practical for QUEEN to be applied in real-world scenario.

\section{Discussion}

The effectiveness of QUEEN is established upon the two main components: sensitivity measurement and output perturbation.
By launching gradient reverse on the sensitive query and feature perturbation on the non-sensitive query, QUEEN achieves the goal where the attack accuracy of the piracy model is significantly lowered while the defense accuracy remaining high.
However, this defense essentially modifies the posterior outputs of the protectee model.
In fact, after receiving the outputs from the defender, the adversary is allowed to take any operation on the paired auxiliary dataset.
This means that if the adversary uses advanced attacks such as D-DAE to purify the perturbed outputs with the generator network, the attack accuracy will be lifted.
QUEEN is capable of defending such advanced attacks to an extent, but the effectiveness is limited if the generator used for purifying the outputs can make the argmax of the softmax output correct.
To mitigate the advanced attacks could be a future research direction.
However, to launch the advanced attacks, the adversary is required to train a large amount of shadow models, meta-classifiers, whose cost is non-negligible.

Currently, the mainstream defenses \cite{lee2018defending, orekondy2019prediction, kariyappa2020defending, tangmodelguard} counter the model extraction attacks based on perturbation of the softmax outputs.
In these studies, the adversary is assumed to train the piracy model with the softmax outputs, because this leads to better test performance.
Using multiple accounts to decrease the defense strength can be countered by IP detection defenses.
A more complex attack is feasible such as querying the model in the distributed denial-of-service (DDOS) manner, but the cost of the attack will significantly increase.

\section{Conclusion}
We propose QUEEN, a proactive defense against model extraction attacks by detecting potential threats of the queries by measuring the cumulative query sensitivity.
The softmax of the query whose feature is within the sensitive region is perturbed to make gradient reverse if the cumulative query sensitivity exceeds the threshold.
The features of the non-sensitive queries are perturbed so as to generate perturbed softmax.
Through extensive experiments, the effectiveness of QUEEN has been proved, where QUEEN is capable of defending the current model extraction attacks and has outperformed the SOTA defenses.
The attack accuracy is decreased at an acceptable cost of the defense accuracy.



%


\bibliographystyle{IEEEtran}
\bibliography{ref}

\appendix

\subsection{Experiment Settings}\label{sect:exp_settings}

\header{Datasets and Model Architectures.}
The information of the training datasets and the corresponding model architectures is summarized in Table \ref{tab:experiment_settings}.
Four types of model architectures are used for training the protectee models on four image classification datasets, which are Caltech256 \cite{griffin_holub_perona_2022}, CUB200\cite{Wah2011TheCB}, CIFAR100, and CIFAR10 \cite{Krizhevsky2009LearningML}.
For the defender, the protectee models are all trained with Outlier Exposure (OE) datasets, because one of the defense methods, Adaptive Misinformation \cite{kariyappa2020defending}, requires it.
For Caltech256 and CUB200, the defender employs ResNet50 \cite{he2016deep}.
For CIFAR100 and CIFAR10, the defender uses VGG16-BN \cite{simonyan2014very}.
The test accuracy of the protectee models is presented in Table \ref{tab:experiment_settings}.
For the adversary, the same model architectures are selected to allow the adversary to obtain the best attack accuracy.
In terms of the auxiliary datasets, the adversary use TinyImageNet200 \cite{tiny-imagenet} for CIFAR10 and CIFAR100, and ImageNet1k \cite{imagenet} for CUB200 and Caltech256.

\begin{table*}[t]
    \caption{Experiment Settings}
    \label{tab:experiment_settings}
    \centering
    \renewcommand\arraystretch{1.}
    \setlength{\tabcolsep}{2.mm}{
    \begin{tabular}{c|ccccc}
        \toprule
        \multirow{4}{*}{Defender} & Training Dataset & CIFAR10 & CIFAR100 & CUB200 & Caltech256 \\
        & OE Dataset & SVHN & SVHN & Indoor67 & Indoor67\\
        & Model Architecture & VGG16-BN & VGG16-BN & ResNet50 & ResNet50\\
        & Accuracy & 92.74\% & 76.12\% & 82.42\% & 86.84\%\\
        \midrule
        \multirow{2}{*}{Adversary} & Auxiliary Dataset & TinyImageNet200 & TinyImageNet200 & ImageNet1k & ImageNet1k \\
        & Model Architecture & VGG16-BN & VGG16-BN & ResNet50 & ResNet50\\
        \bottomrule
    \end{tabular}}
\end{table*}

\header{Evaluation Metric.}
In the experiment, we mainly use attack accuracy and attack agreement as the evaluation metrics for the attack performance.
Attack accuracy is defined as the ratio of the correctly classified samples to the total samples in the test dataset of the protectee model.
Attack agreement, on the other hand, is the ratio of the samples that are identically classified by both the protectee and piracy models to the total samples in the test dataset.
The lower the values of the two metrics are, the better the defense is.

\header{Implementation of QUEEN}
For the mapping network, we use four fully connected layers to map the training features to the 2D space.
As for the piracy model ensemble, we select ResNet18 for Caltech256 and CUB200 datasets, and VGG11-BN for the CIFAR-10 and CIFAR-100 datasets.
The mapping network is trained for $100$ epochs using the SGD optimizer, where the learning rate is set to $0.01$ and it decreases by $0.5$ every $20$ epochs.
The members of the piracy model ensemble are trained for $5$ 
epochs on the sub-datasets in each of which there are $500$ samples in each class. 

\begin{table}[t!]
    \centering
    \caption{Impact of the Number of Shadow Models on the Attack Accuracy of the Piracy Models on CIFAR10 Attacked by pBayes.}\label{tab:num_shadow_cifar10_pbayes}
    \setlength{\tabcolsep}{1.5mm}{
    \begin{tabular}{c|c|c|c|c|c}
    \toprule
    Number of Models & 1 & 2 & 3 & 5 & 10 \\
    \midrule
    Attack Accuracy & 67.12\% & 66.41\% & 65.84\% & 65.33\% & 65.67\%\\
    Attack Agreement & 69.41\% & 68.57\% & 67.76\% & 67.56\% & 67.81\%\\
    \bottomrule
    \end{tabular}}
\end{table}

\begin{table}[t!]
    \centering
    \caption{Impact of the Number of Shadow Models on the Attack Accuracy of the Piracy Models on CIFAR100 Attacked by D-DAE+.}\label{tab:num_shadow_cifar100_ddaeplus}
    \setlength{\tabcolsep}{1.5mm}{
    \begin{tabular}{c|c|c|c|c|c}
    \toprule
    Number of Models & 1 & 2 & 3 & 5 & 10 \\
    \midrule
    Attack Accuracy & 57.57\% & 56.78\% & 57.12\% & 56.59\% & 56.61\%\\
    Attack Agreement & 59.46\% & 58.89\% & 59.25\% & 59.11\% & 58.57\%\\
    \bottomrule
    \end{tabular}}
\end{table}

\begin{table}[t!]
    \centering
    \caption{Impact of the Number of Shadow Models on the Attack Accuracy of the Piracy Models on CIFAR100 Attacked by pBayes.}\label{tab:num_shadow_cifar100_pbayes}
    \setlength{\tabcolsep}{1.5mm}{
    \begin{tabular}{c|c|c|c|c|c}
    \toprule
    Number of Models & 1 & 2 & 3 & 5 & 10 \\
    \midrule
    Attack Accuracy & 57.32\% & 56.51\% & 56.72\% & 56.84\% & 56.45\%\\
    Attack Agreement & 59.81\% & 59.04\% & 59.31\% & 58.95\% & 58.67\%\\
    \bottomrule
    \end{tabular}}
\end{table}

\begin{figure}[t!]
    \centering
    \includegraphics[width=.4\textwidth]{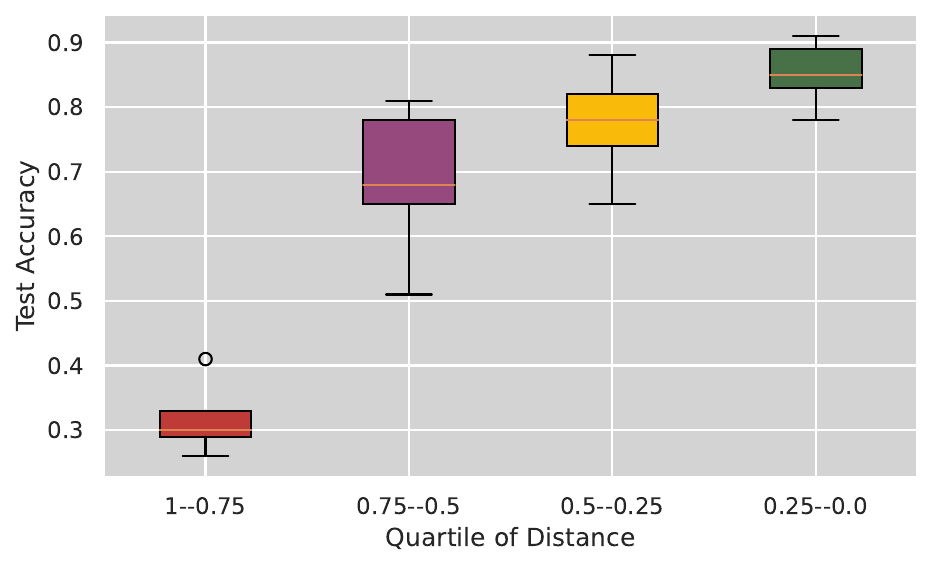}
    \caption{Accuracy of Models Trained on Peripheral and Central Data.}
    \label{fig:ablation_peri_vs_cent}
\end{figure}

\subsection{Peripheral Data vs. Central Data.}
\label{appdendix:peri_vs_central}
We split the training dataset of MNIST to train models and test their accuracy.
The cluster center of each class is computed such that the samples in each class is ranked based on the distance between their features and the cluster centers.
The closer the query feature to the cluster center is, the lower the rank of the feature will be.
We split the training dataset into four quartiles based on the rank.
For example, the $0.25$--$0.0$ quartile is the subset containing the top $25\%$ of the training samples that are closest to the cluster centers.
In each training attempt, $500$ samples are randomly sampled from each class of the quartile, and then used for training the model.
For each quartile, we repeat for five times to get the results, where the settings remain the same as in training the protectee models on MNIST in the previous experiments.

As depicted in Figure \ref{fig:ablation_peri_vs_cent}, the results support our assumption, where the peripheral subset leads to significantly lower test accuracy than the central subset.
This suggests that the central data samples are more valuable than the peripheral data samples in terms of protection.
Output perturbation that will lower the defense accuracy such as gradient reverse should be applied to the central data, because the attack accuracy can be efficiently lowered at a less cost of defense accuracy.
This explains the idea of QUEEN: punishing the critical queries and letting the trivial queries pass.

\begin{figure}[t]
    \centering
    \includegraphics[width=.45\textwidth]{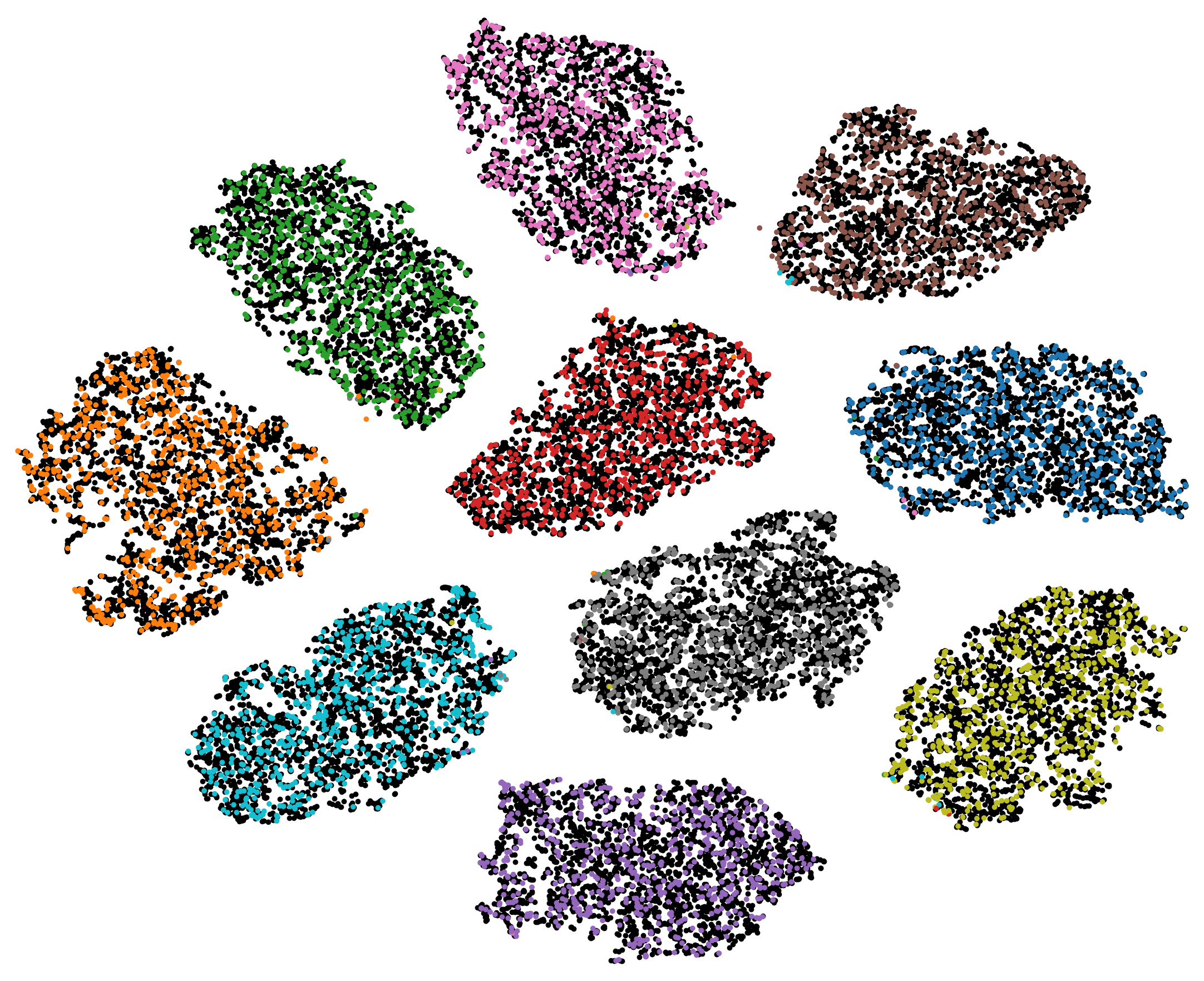}
    \caption{Visualization of the decision boundary via t-SNE. The black/colorful dots denote the features of the training/test data, where each color represents one class.}
    \label{fig:tsne_decision_boundary}
\end{figure}

\subsection{Supplementary of Theoretical Analysis}\label{sect:supp_theo_als}

\subsubsection{Proof of Theorem 1}
\label{appendix:gradient_reverse}

\header{Theorem 1.}
\textit{(The sufficient condition of gradient reverse) To achieve gradient reverse in Definition 4, it is enough to set  $\ybt = 2\ybh' - \ybh$.}

\header{Proof.}
We consider the widely used cross-entropy loss here $\mathcal L_{CE}=-\sum_{i=1}^n \y^i\log(\ybh^i)$, where $\y^i$ is the $i$-th element of the ground-truth label vector $\y$.

According to the chain rule,
\begin{equation}
    \nabla_{\tau}\mathcal L_{CE}=\frac{\partial\mathcal L_{CE}}{\partial \ybh'}\cdot\frac{\partial \ybh'}{\partial h}\cdot\frac{\partial h}{\partial \tau},
\end{equation}
where $\ybh' = \sm(h(\x'))$.

The gradient reverse in Definition 4 requires that
\begin{equation}
    \frac{\partial\mathcal L_{CE}}{\partial \ybt} \cdot\frac{\partial \ybt}{\partial h} \cdot\frac{\partial h}{\partial \tau} = -\frac{\partial\mathcal L_{CE}}{\partial \ybh'}\cdot\frac{\partial \ybh'}{\partial h} \cdot\frac{\partial h}{\partial \tau}.
\end{equation}

Eventually, we have
\begin{equation}
    \ybt = 2 \ybh' - \ybh.
\end{equation}
That completes the proof.\hfill$\square$

\subsubsection{Proof of Theorem 2}
\label{appendix:certifiability}

\header{Theorem 2.}
\textit{Given a learning algorithm that learns a piracy model $h$, let $\eta$ be the actual number of honestly answered sensitive queries, the piracy model can at most be trained to have the maximum allowable error $\epsilon$ and the upperbound of error probability $\delta$, if}
\begin{equation}
\begin{aligned}
    \eta \leq \hat{\eta} = \frac{1}{2\epsilon^2}\cdot\ln(\frac{2}{\delta}),
\end{aligned}
\end{equation}
\textit{where $\hat{\eta}$ is the maximum number of honestly answered sensitive queries.}
\textit{Further, let $t$ be the threshold, and $r$ be the query radius, we will have the relationship as follows.}
\begin{equation}
\begin{aligned}
    r \geq \sqrt{\frac{2t}{\ln(\frac{2}{\delta})}}\epsilon\bd,
\end{aligned}
\end{equation}
\textit{where $\bar d$ is a constant denoting the average distance of features to their cluster center.}

\header{Proof.}
According to hoeffding's inequality, we have
\begin{equation}
\begin{aligned}
    Pr\bigg(|E(h) - E(h_e)| > \epsilon \bigg) \leq 2e^{-2\hat{\eta}\epsilon^2}.
\end{aligned}
\end{equation}
Let $\delta = 2e^{-2\hat{\eta}\epsilon^2}$, we then have
\begin{equation}\label{eq:28}
\begin{aligned}
    \hat{\eta} = \frac{1}{2\epsilon^2}\cdot\ln(\frac{2}{\delta}).
\end{aligned}
\end{equation}
Thus, the piracy model can at most achieve $\epsilon$ error rate with $\delta$ probability if $\eta \leq \hat{\eta}$.

So far, we have determined the maximum number of honestly answered queries $\eta$ by $\epsilon$ and $\delta$.

In addition, $\eta$ can be estimated by $t$ and $r$:
\begin{equation}\label{eq:29}
\begin{aligned}
    \eta = \frac{t\pi\bd^2}{\pi r^2} = \frac{t\bd^2}{r^2}
\end{aligned}
\end{equation}
Combine equations \ref{eq:28} and \ref{eq:29} with the condition $\eta \leq \hat{\eta}$, we have
\begin{equation}
\begin{aligned}
    \frac{t\bd^2}{r^2} \leq \frac{1}{2\epsilon^2}\cdot\ln(\frac{2}{\delta})
\end{aligned}
\end{equation}
Eventually, we have
\begin{equation}
\begin{aligned}
    r \geq & \sqrt{\frac{2t}{\ln(\frac{2}{\delta})}}\epsilon\bd
\end{aligned}
\end{equation}
This completes the proof.\hfill$\square$

\subsubsection{Proof of Theorem 3}

\header{Theorem 3.} \cite{tangmodelguard,cover1999elementsOfInfoTheory}
\textit{
Any adaptive model extraction attack with an arbitrary recovery function $R(\cdot)$ cannot attain a smaller gap between the recovered predictions $R(\tilde{Y}) \in \mbb R^{M\times N}$ and the original predictions $Y \in \mbb R^{M \times N}$ than the following lower bound:
}
\begin{equation}
\begin{aligned}
    \mbb E[\|R(\tilde{Y}) - Y\|^2_2] \geq \frac{MN}{2\pi e} \exp{\bigg(\frac{2}{MN} h(Y|\tilde{Y}) \bigg)},
\end{aligned}
\end{equation}
\textit{
where $M$ and $N$ respectively denote the number of the samples and classes;
$h(Y|\tilde{Y})$ is the conditional entropy.
}

\textit{Proof.}
The following inequality holds for $\y$ with an arbitrary distribution conditioned on the event $\{\tilde{Y} = \tilde{Y}_p\}$ \cite{cover1999elementsOfInfoTheory}:
\begin{equation}
\begin{aligned}
    h(Y|\tilde{Y}=\tilde{Y}_p) \leq \frac{1}{2}\log((2\pi e)^{MN} \det(\text{Cov}(Y|\tilde{Y}=\tilde{Y}_p)))\\
    \Rightarrow \det(\text{Cov}(Y|\tilde{Y}=\tilde{Y}_p))) \geq \frac{1}{2\pi e)^{MN}} \exp(2h(Y|\tilde{Y}=\tilde{Y}_p))
\end{aligned}
\end{equation}
where the equality holds with Gaussian $Y|\{\tilde{Y}=\tilde{Y}_p\}$; $\tilde{Y}_p$ denotes the perturbed outputs generated by the perturbation function given the training labels.
With the fact that any recovery function $R(\cdot)$ tries to minimize $\mbb E[\|R(\tilde{Y} - Y)\|^2_2|\tilde{Y}]$, we then have:
\begin{equation}
\begin{aligned}
    \mbb E & [\|R(\tilde{Y}) - Y\|^2_2|\tilde{Y}=\tilde{Y}_p]\\
    & \geq \mbb E[\|Y - \mbb E[Y|\tilde{Y}=\tilde{Y}_p] \|^2_2|\tilde{Y}=\tilde{Y}_p]\\
    & =\text{tr}(\text{Cov}(Y|\tilde{Y}=\tilde{Y}_p))\\
    & \geq NC[\det(\text{Cov}(Y|\tilde{Y}=\tilde{Y}_p))]^{\frac{1}{MN}} \\
    & \geq \frac{MN}{2\pi e} \exp \bigg(\frac{2}{MN}h(Y|\tilde{Y}=\tilde{Y}_p) \bigg).
\end{aligned}
\end{equation}
That completes the proof.\hfill$\square$

\begin{table*}[t!]
    \caption{Evaluation of defenses against attacks on CIFAR100.}
    \label{tab:eval_cifar100}
    \centering
    \begin{tabular}{ccccccccccc}
        \toprule
        Query Method & Attack Method & None & RS & MAD & AM & Label-only & Rounding & EMDP & ModelGuard & QUEEN \\
        \midrule
        \multirow{7}{*}{KnockoffNet} & Direct Query & 66.74\% & 62.55\% & 58.79\% & 60.75\% & 54.47\% & 62.91\% & 48.14\% & 51.98\% & \textbf{1.0\%}\\
         & Label-Only & 54.47\% & 54.47\% & 54.47\% & \textbf{52.61\%} & 54.47\% & 54.47\% & 54.47\% & 54.47\% & 53.47\%\\
         & S4L & 63.21\% & 57.14\% & 55.39\% & 60.13\% & 55.18\% & 61.75\% & 50.18\% & 46.42\% & \textbf{1.0\%}\\
         & Smoothing & 64.27\% & 63.41\% & 60.17\% & 62.53\% & 61.43\% & 66.18\% & 55.81\% & 51.84\% & \textbf{1.0\%}\\
         & D-DAE & 66.74\% & 63.19\% & 62.51\% & 61.34\% & 58.61\% & 63.31\% & 62.46\% & 59.10\% & \textbf{58.37\%}\\
         & D-DAE+ & 66.74\% & 64.77\% & 64.37\% & 62.10\% & 58.91\% & 63.17\% & 62.66\% & 57.01\% & \textbf{56.61\%}\\
         & pBayes & 66.74\% & 65.55\% & 65.01\% & 64.97\% & 58.18\% & 65.47\% & 58.41\% & 59.11\% & \textbf{56.45\%}\\
         \hline
         \multirow{5}{*}{JBDA-TR} & Direct Query & 41.41\% & 35.17\% & 13.35\% & 30.61\% & 22.32\% & 36.11\% & 9.43\% & 5.87\% & \textbf{1.0\%}\\
         & Label-Only & 22.32\% & 22.32\% & 22.32\% & \textbf{15.32\%} & 22.32\% & 22.32\% & 22.32\% & 22.32\% & 20.36\%\\
         & D-DAE & 41.41\% & 30.18\% & 20.54\% & 26.63\% & 19.78\% & 24.71\% & 24.76\% & 19.31\% & \textbf{18.36\%}\\
         & D-DAE+ & 41.41\% & 38.87\% & 22.24\% & 31.44\% & 21.68\% & 39.96\% & 29.14\% & 20.62\% & \textbf{17.18\%}\\
         & pBayes & 41.41\% & 39.52\% & 27.46\% & 40.04\% & 24.11\% & 39.75\% & 26.91\% & 22.61\% & \textbf{20.75\%}\\
         \hline
         \multicolumn{2}{c}{Max Piracy Model Accuracy} & 66.74\% & 65.55\% & 65.01\% & 64.97\% & 61.43\% & 66.18\% & 62.66\% & 59.11\% & \textbf{58.37\%}\\
         \multicolumn{2}{c}{Max Piracy Model Agreement} & 72.57\% & 71.15\% & 71.24\% & 71.48\% & 66.91\% & 68.41\% & 64.28\% & 65.82\% & \textbf{65.51\%}\\
         \multicolumn{2}{c}{Protectee Model Accuracy} & 76.12\% & 76.12\% & 76.12\% & 74.45\% & 76.12\% & 76.12\% & 76.12\% & 76.12\% & 74.23\%\\
         \bottomrule
    \end{tabular}
\end{table*}

\begin{table*}[t!]
    \caption{Evaluation of defenses against attacks on CUB200.}
    \label{tab:eval_cub200}
    \centering
    \begin{tabular}{ccccccccccc}
        \toprule
        Query Method & Attack Method & None & RS & MAD & AM & Label-only & Rounding & EMDP & ModelGuard & QUEEN \\
        \midrule
        \multirow{7}{*}{KnockoffNet} & Direct Query & 80.79\% & 75.41\% & 67.15\% & 77.14\% & 73.11\% & 79.91\% & 71.47\% & 54.02\% & \textbf{0.5\%}\\
         & Label-Only & 73.11\% & 73.11\% & 73.11\% & \textbf{68.15\%} & 73.11\% & 73.11\% & 73.11\% & 73.11\% & 69.88\%\\
         & S4L & 80.14\% & 74.76\% & 60.55\% & 75.84\% & 76.51\% & 78.47\% & 72.78\% & 54.15\% & \textbf{0.5\%}\\
         & Smoothing & 79.91\% & 75.04\% & 67.11\% & 74.52\% & 75.51\% & 79.35\% & 75.88\% & 51.35\% & \textbf{0.5\%}\\
         & D-DAE & 80.79\% & 78.35\% & 78.76\% & 76.49\% & 72.18\% & 79.81\% & 76.98\% & \textbf{65.47\%} & 70.19\%\\
         & D-DAE+ & 80.79\% & 79.31\% & 79.89\% & 77.40\% & 71.83\% & 79.61\% & 77.21\% & 74.26\% & \textbf{68.34\%}\\
         & pBayes & 80.79\% & 80.28\% & 79.38\% & 78.81\% & 70.58\% & 80.31\% & 77.57\% & 76.47\% & \textbf{69.14\%}\\
         \hline
         \multirow{5}{*}{JBDA-TR} & Direct Query & 64.23\% & 54.15\% & 10.14\% & 36.17\% & 30.66\% & 50.91\% & 14.22\% & 5.14\% & \textbf{0.5\%}\\
         & Label-Only & 30.66\% & 30.66\% & 30.66\% & 23.70\% & 30.66\% & 30.66\% & 30.66\% & 30.66\% & \textbf{23.34\%}\\
         & D-DAE & 64.23\% & 53.25\% & 16.73\% & 38.19\% & 25.72\% & 44.29\% & 32.91\% & \textbf{8.91\%} & 15.81\%\\
         & D-DAE+ & 64.23\% & 62.78\% & 34.79\% & 41.74\% & 33.18\% & 48.85\% & 36.48\% & 28.37\% & \textbf{17.21\%}\\
         & pBayes & 64.23\% & 62.39\% & 34.74\% & 61.69\% & 32.43\% & 50.03\% & 33.17\% & 27.17\% & \textbf{18.55\%}\\
         \hline
         \multicolumn{2}{c}{Max Piracy Model Accuracy} & 80.79\% & 80.28\% & 79.89\% & 78.81\% & 76.51\% & 80.31\% & 77.57\% & 76.47\% & \textbf{70.19\%}\\
         \multicolumn{2}{c}{Max Piracy Model Agreement} & 84.21\% & 84.02\% & 83.57\% & 80.28\% & 79.04\% & 81.35\% & 78.94\% & 78.21\% & \textbf{73.47\%}\\
         \multicolumn{2}{c}{Protectee Model Accuracy} & 82.42\% & 82.42\% & 82.42\% & 80.19\% & 82.42\% & 82.42\% & 82.42 & 82.42\% & 80.71\%\\
         \bottomrule
    \end{tabular}
\end{table*}

\begin{table*}[t!]
    \caption{Evaluation of defenses against attacks on Caltech256.}
    \label{tab:eval_caltech256}
    \centering
    \begin{tabular}{ccccccccccc}
        \toprule
        Query Method & Attack Method & None & RS & MAD & AM & Label-only & Rounding & EMDP & ModelGuard & QUEEN \\
        \midrule
        \multirow{7}{*}{KnockoffNet} & Direct Query & 83.57\% & 77.24\% & 67.21\% & 75.63\% & 72.98\% & 78.81\% & 69.74\% & 53.37\% & \textbf{0.39\%}\\
         & Label-Only & 72.98\% & 72.98\% & 72.98\% & \textbf{68.33\%} & 72.98\% & 72.98\% & 72.98\% & 72.98\% & 69.41\%\\
         & S4L & 80.43\% & 75.86\% & 62.01\% & 75.82\% & 72.17\% & 77.49\% & 68.41\% & 54.73\% & \textbf{0.39\%}\\
         & Smoothing & 81.77\% & 75.32\% & 69.14\% & 73.55\% & 76.04\% & 77.41\% & 75.93\% & 55.76\% & \textbf{0.39\%}\\
         & D-DAE & 83.57\% & 80.25\% & 79.76\% & 78.35\% & 72.68\% & 75.04\% & 73.92\% & 68.31\% & \textbf{66.26\%}\\
         & D-DAE+ & 83.57\% & 81.17\% & 81.43\% & 78.65\% & 70.33\% & 79.17\% & 73.21\% & 70.65\% & \textbf{66.45\%}\\
         & pBayes & 83.57\% & 81.75\% & 82.21\% & 82.65\% & 72.65\% & 82.43\% & 76.14\% & 78.34\% & \textbf{72.44\%}\\
         \hline
         \multirow{5}{*}{JBDA-TR} & Direct Query & 64.69\% & 54.51\% & 9.17\% & 38.92\% & 31.25\% & 44.21\% & 12.54\% & 5.18\% & \textbf{0.39\%}\\
         & Label-Only & 28.91\% & 28.91\% & 28.91\% & 24.91\% & 28.91\% & 28.91\% & 28.91\% & 28.91\% & \textbf{23.15\%}\\
         & D-DAE & 64.69\% & 55.57\% & 25.79\% & 37.21\% & 29.31\% & 45.19\% & 48.17\% & \textbf{15.26\%} & 29.76\%\\
         & D-DAE+ & 64.69\% & 62.41\% & 38.79\% & 44.99\% & 38.08\% & 55.13\% & 49.56\% & 37.76\% & \textbf{37.45\%}\\
         & pBayes & 64.69\% & 61.21\% & 38.77\% & 59.74\% & 36.71\% & 58.62\% & 41.73\% & 35.13\% & \textbf{32.51\%}\\
         \hline
         \multicolumn{2}{c}{Max Piracy Model Accuracy} & 83.57\% & 81.75\% & 82.21\% & 82.65\% & 76.04\% & 82.43\% & 76.14\% & 78.34\% & \textbf{72.44\%}\\
         \multicolumn{2}{c}{Max Piracy Model Agreement} & 86.14\% & 85.89\% & 86.01\% & 85.13\% & 79.14\% & 85.22\% & 79.34\% & 83.02\% & \textbf{75.63\%}\\
         \multicolumn{2}{c}{Protectee Model Accuracy} & 86.84\% & 86.84\% & 86.84\% & 84.24\% & 86.84\% & 86.84\% & 86.84\% & 86.84\% & 83.16\%\\
         \bottomrule
    \end{tabular}
\end{table*}

\begin{table*}[t!]
    \caption{Time of processing $1,000$ queries using each defense on CIFAR10.}
    \label{tab:query_time_cifar10}
    \centering
    \begin{tabular}{c|cccccccccc}
        \toprule
        Defenses & None & RS & MAD & AM & Label-only & Rounding & EMDP & ModelGuard & QUEEN \\
        \midrule
        Time (Seconds) & 1.28 & 1.56 & 30.76 & 2.37 & 1.21 & 1.37 & 1.52 & 2.47 & 4.28\\
        \bottomrule
    \end{tabular}
\end{table*}




\end{document}